\documentclass[twocolumn,showpacs,preprintnumbers,amsmath,amssymb]{revtex4}
\usepackage{tabularx,graphicx,subfigure}\begin{document}
\newcommand{\beq}{\begin{equation}}
\newcommand{\eeq}{\end{equation}}
\newcommand{\beqn}{\begin{eqnarray}}
\newcommand{\eeqn}{\end{eqnarray}}
\newcommand{\bmath}{\begin{subequations}}
\newcommand{\emath}{\end{subequations}}

\newcommand{\up}{\uparrow}
\newcommand{\dow}{\downarrow}

\newcommand{\kf}{k_{\scriptscriptstyle{F}}}

\newcommand{\bfk}{\boldsymbol{k}}
\newcommand{\bfq}{\boldsymbol{q}}
\newcommand{\RSS}{\emph{RSS}}

\newcommand{\cdg}{c^{\scriptscriptstyle{\dagger}}_{\scriptstyle{\boldsymbol{k} \eta}}}
\newcommand{\cdo}{c^{\scriptscriptstyle{\dagger}}_{\scriptstyle{\boldsymbol{k} \om}}}

\newcommand{\cnd}{c_{\scriptstyle{\boldsymbol{k} \eta}}}
\newcommand{\cno}{c_{\scriptstyle{\boldsymbol{k} \om}}}
\newcommand{\cds}{c^{\scriptscriptstyle{\dagger}}_{\scriptstyle{\boldsymbol{k} \sigma}}}
\newcommand{\cns}{c_{\scriptstyle{\boldsymbol{k} \sigma'}}}
\newcommand{\cdj}{c^{\scriptscriptstyle{\dagger}}_{\scriptstyle{j \sigma}}}
\newcommand{\cnj}{c_{\scriptstyle{j' \sigma'}}}
\newcommand{\cda}{c^\dagger}
\newcommand{\tki}{\tilde{k}_i}
\newcommand{\tkj}{\tilde{k}'_j}
\newcommand{\cdp}{c^\dagger_{\boldsymbol{k}+}}
\newcommand{\cdm}{c^\dagger_{\boldsymbol{k}-}}
\newcommand{\cnp}{c_{\boldsymbol{k} +}}
\newcommand{\cnm}{c_{\boldsymbol{k} -}}
\newcommand{\bpm}{\begin{pmatrix}}
\newcommand{\epm}{\end{pmatrix}}
\newcommand{\nk}{n_{\boldsymbol{k}'}}
\newcommand{\dk}{\Delta_{\boldsymbol{k}'}}
\newcommand{\Dk}{\Delta_{\boldsymbol{k}}}
\newcommand{\tk}{\tilde{k}}
\newcommand{\tkp}{\tilde{k}'}
\newcommand{\bal}{\begin{align}}
\newcommand{\eal}{\end{align}}
\newcommand{\A}{\mathrm{\AA^{\scriptscriptstyle{-1}}}}
\newcommand{\R}{\alpha_{\scriptscriptstyle{R}}}
\newcommand{\V}{V_{\text{eff}}}
\newcommand{\si}{\text{sin}\,}
\newcommand{\co}{\text{\text{cos}}\,}
\newcommand{\te}{t_{\text{eff}}}
\newcommand{\ep}{\bar{\epsilon}}

\newcommand{\efu}{\varepsilon_{\scriptscriptstyle{F}}^{\scriptscriptstyle{\uparrow}}}
\newcommand{\efd}{\varepsilon_{\scriptscriptstyle{F}}^{\scriptscriptstyle{\downarrow}}}
\newcommand{\ef}{\varepsilon_{\scriptscriptstyle{F}}}

\newcommand{\ab}{\alpha^2 + \beta^2}
\newcommand{\ag}{\alpha^2 + \gamma^2}

\newcommand{\om}{\scriptscriptstyle{\Omega}}

\title{Effect of Electron-Electron Interactions on Rashba-like and Spin-Split Systems}
\author{A. Alexandradinata, J. E. Hirsch }
\address{Department of Physics, University of California, San Diego,
La Jolla, CA 92093-0319}

\begin{abstract} 
The role of electron-electron interactions is analyzed for Rashba-like and spin-split systems within a tight-binding single-band Hubbard model with on-site and all nearest-neighbor matrix elements of the Coulomb interaction. By Rashba-like systems we refer to the Dresselhaus and Rashba spin-orbit coupled phases, while spin-split systems have spin-up and spin-down Fermi surfaces shifted relative to each other. Both systems break parity but preserve time-reversal symmetry. They belong to a class of symmetry-breaking ground states that satisfy: (i) electron crystal momentum is a good quantum number (ii) these states have no net magnetic moment and (iii) their distribution of `polarized spin' in momentum space breaks the lattice symmetry. For all members of this class, the relevant Coulomb matrix elements are found to be nearest-neighbor exchange $J$, pair-hopping $J'$ and nearest-neighbor repulsion $V$. These ground states lower their energy most effectively through $J$, hence we name them Class $J$ states. The competing effects of $V-J'$ on the direct and exchange energies determine the relative stability of Class $J$ states. We show that the spin-split and Rashba-like phases are the most favored ground states within Class $J$ because they have the minimum anisotropy in `polarized spin'. We analyze these two states on a square lattice and find that the spin-split phase is always favored for near-empty bands; above a critical filling, we predict a transition from the paramagnetic to the Rashba-like phase at a critical $J\,(J_{c1})$ and a second transition from the Rashba-like to the spin-split state at $J_{c2}>J_{c1}$. An energetic comparison with ferromagnetism highlights the importance of the role of $V$ in the stability of Class $J$ states. We discuss the relevance of our results to (i) the $\alpha$ and $\beta$ phases proposed by Wu and Zhang in the Fermi Liquid formalism and (ii) experimental observations of spin-orbit splitting in \emph{Au}(111) surface states. 
\end{abstract}
\maketitle

\section{introduction}

Numerous broken-symmetry phases have been predicted to arise from electron-electron interactions. These phases are often framed in the language of Pomeranchuk instabilities of the Fermi liquid \cite{pom,gorkov,fradkin,metzner}. In this language, the interactions that drive a symmetry breaking are purely phenomenological, though work has been done to derive the Landau interaction parameters from a more physical model, e.g. the Hubbard model; in these works \cite{Yuki, Frigeri} particular attention has been drawn to the Hubbard parameter $U$ (describing on-site repulsion) because of the strong interest in high-temperature d-wave superconductivity \cite{Halboth}. There has been further progress made by Lamas and coworkers in understanding how the lattice affects FL instabilities in two dimensions \cite{Lamas}, focusing on Hubbard interactions $U$ and $V$ (nearest-neighbor repulsion).\\
 
In the tight-binding representation, $U$ and $V$  are matrix elements of the Coulomb interaction that are diagonal in the density operators; there are also non-diagonal matrix elements \cite{hubbard, kssh, ph,scfm,fm}. In total, there are four nearest-neighbor interactions, denoted by $V$ (diagonal) and $J$, $J'$ and $\Delta t$ (off-diagonal). This Hamiltonian (including $U$) is usually termed `generalized Hubbard model'.\\


In this paper, we ask: other than ferromagnetism and antiferromagnetism, what kinds of broken-symmetry phases involving spin arise in the generalized Hubbard model? It was initially proposed by one of us \cite{hirschss} that the interaction  $J$ (nearest-neighbor ferromagnetic exchange) plays a crucial role in displacing the Fermi surfaces of spin up and down electrons relative to one another; the resultant spin-split metallic phase is illustrated in Fig. (1-a). Subsequently, the spin-split phase has also been predicted to arise in the FL framework by Wu and coworkers \cite{wu} and by Chubukov and Maslov \cite{chubukov}. A possible realization of the spin-split phase was originally proposed to be the low-temperature broken-symmetry phase of Chromium \cite{hirschss}; recently Varma and Zhu have proposed that it may describe the `hidden order' phase in the heavy fermion compound $U Ru_2 Si_2$ \cite{varma}.\\

In addition to the spin-split phase (dubbed the $\alpha$ phase by Wu et al), Wu et al also showed that many spin-orbit coupled phases (the $\beta$ phase) may arise from a phenomelogical Landau interaction in the p-wave spin channel. In two dimensions, these phases include the Rashba \cite{rashba}, Dresselhaus \cite{dress} and helicity spin-orbit coupled systems; for convenience, we call these group of spin-orbit coupled phases Rashba-like. The Rashba-polarized state is illustrated in Fig. (1-b). \\

\begin{figure}[h]
 \begin{center}
		\subfigure[]{\resizebox{7cm}{!}{\includegraphics[width=6cm, height=6cm]{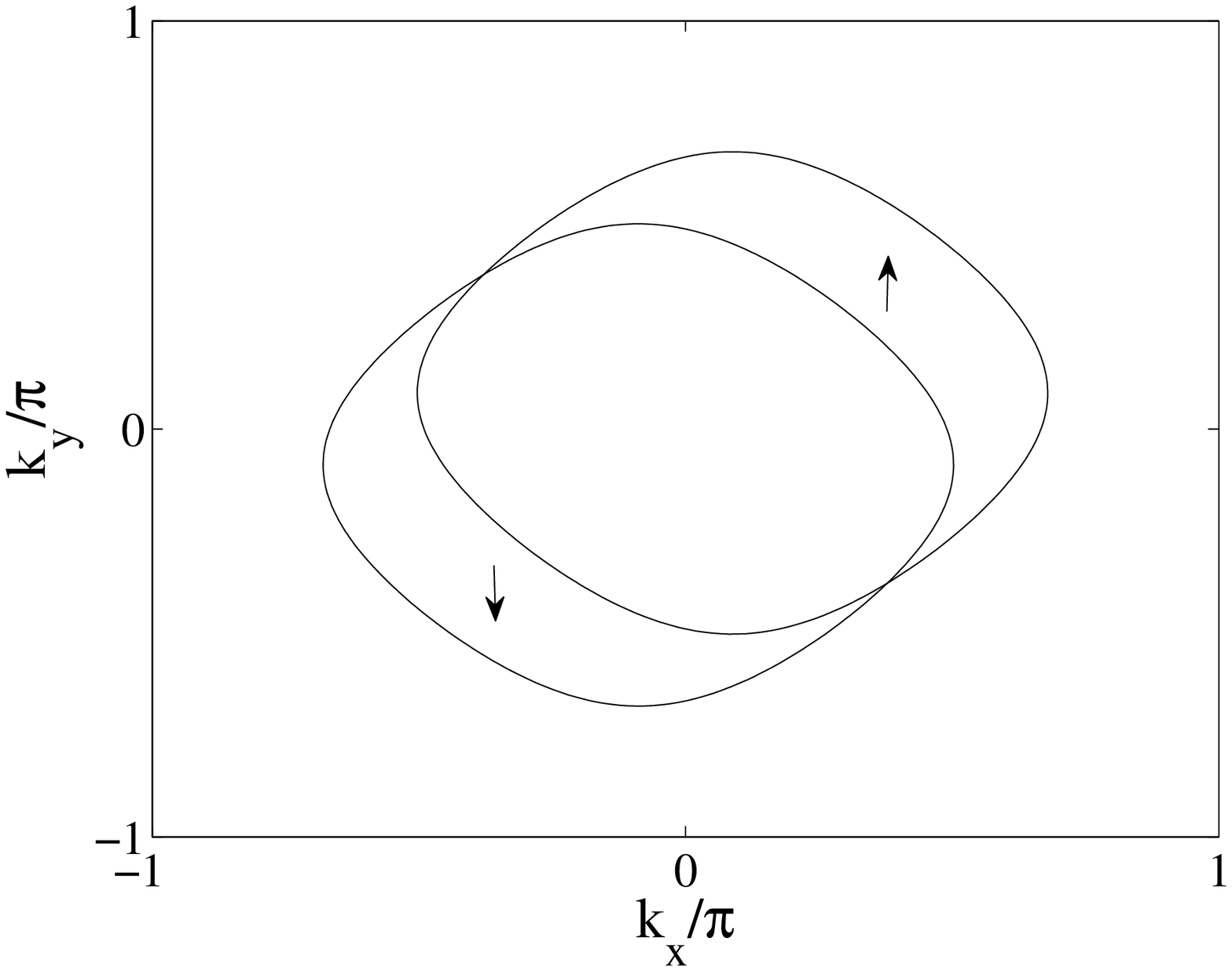}}}
		\subfigure[]{\resizebox{7cm}{!}{\includegraphics[width=6cm, height=6cm]{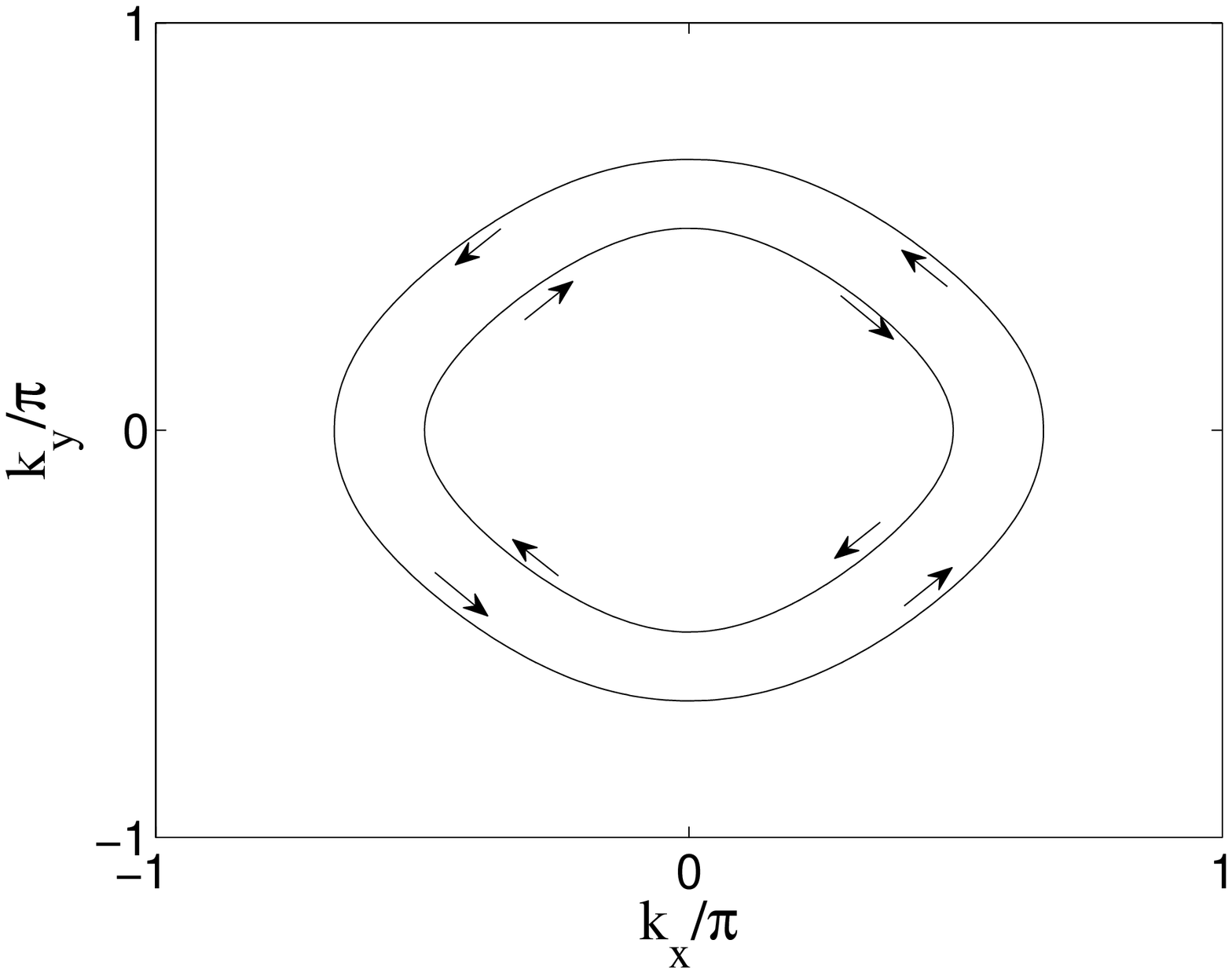}}}
		\end{center}
		\caption{(a) \, Fermi energy contours of a  spin-split state (b) \, Fermi energy contours of a Rashba-polarized state }
\end{figure}

In this paper we return to the tight-binding Hubbard formalism and analyze the effects of on-site repulsion and all nearest-neighbor matrix elements of the Coulomb interaction. We demonstrate that the spin-split state and the Rashba-like phases are part of a broader class of symmetry-breaking ground states for which the relevant matrix elements are nearest-neighbor exchange $J$, pair hopping $J'$ and nearest-neighbor repulsion $V$. This class of ground states has three properties: (i) electron crystal momentum is a good quantum number; (ii) these states have no net magnetic moment; (iii) their distribution of `polarized spin' in momentum space breaks the lattice symmetry. In this class of ground states, we compare the stability of different phases and examine the individual roles played by $J,J'$ and $V$ in changing the dispersion and the ground state energy. \\

In our analysis of the Rashba-polarized state, we are also motivated by the following observation. Photoemission experiments on $Au(111)$ surfaces have revealed a splitting of Rashba-polarized Fermi surfaces; the experimental results may be reproduced in a free-electron model by a phenomelogical interaction 
\bal \label{rash free hami}
H_{R}=\R \vec{\sigma}\cdot(\hat{n}\times\vec{k})
\end{align}
with an electric field perpendicular to the surface (parallel to $\hat{n}$)\cite{au1,au2,au3,au4,au5}. The Fermi wavevector in $Au$ is $k_{\scriptscriptstyle{F}}\sim 0.16 \, \A$ and the measured energy splitting of the Fermi surfaces is $\sim 0.11 \text{eV}$, corresponding to a Rashba coupling constant $\R \sim 0.34 \, \text{eV} \A$. The Rashba coupling constant derived from the Dirac Hamiltonian in the non-relativistic limit is
\bal \label{rash cons dira}
\R=-\frac{e \hbar^2}{4m_e^2c^2} E
\end{align}
with $e$ and $m_e$ the electron charge and mass and $E$ the electric field normal to the surface. For the experimentally observed splitting, the magnitude of the electric field in Eq. (\ref{rash cons dira}) is
\bal \label{elec fiel}
-eE=91,228 \text{eV}/ \A.
\end{align} 
Such a large value of the electric field is justified only with the assumption that the conduction electrons get very close to the $Au$ nucleus; this is incompatible with a free-electron model. In this paper we ask the question: can electron-electron interactions cause a large enhancement of the single-particle spin-orbit splitting so that the observed Fermi surface splitting is explained with a much weaker electric field than the value in Eq. (\ref{elec fiel})?\\
   
This paper is organized as follows. In Sec. II, we discuss the relevance of various Hubbard parameters and identify a class of ground states for which the relevant parameters are $J,J'$ and $V$. With these important parameters, we derive a reduced Hamiltonian. We solve this reduced Hamiltonian for Rashba-like phases in Sec. III and for the spin-split state in Sec. IV. In Sec. V, we find that $V$ plays a pivotal role in the phase transition between the Rashba-like and the spin-split state. In Sec. VI, we consider the stability of Class $J$ ground states other than the spin-split and Rashba-like phases; we also discuss various symmetries of the class. In Sec. VII, we investigate the stability of the spin-split and Rashba-like states in comparison with ferromagnetism. Sec. VIII draws the connection between the formalism in this paper and the Fermi liquid $\alpha$ and $\beta$ phases\cite{wu}. In Sec. IX, we summarize the key physical ideas and discuss possible generalizations to other lattices. Several technical points are discussed in the Appendix. \\

\section{Reduced Hamiltonian}

We consider a single band generalized Hubbard model with on-site and all nearest-neighbor interactions \cite{amadon}:
\bal \label{hami}
H = H_t + H_U + H_J + H_{J'} + H_V + H_{\Delta t}.
\end{align}
Respectively, the individual terms are the nearest-neighbor hopping and the various electron-electron interactions that correspond to on-site repulsion $U$, nearest-neighbor exchange $J$, pair-hopping $J'$, nearest-neighbor repulsion $V$ and correlated hopping $\Delta t$. All the 
electron-electron interaction parameters are positive\cite{scfm}.\\

We consider a two-dimensional square lattice and assume that the spinor is confined in the x-y plane; we denote this restricted set of bases by the index $\{\Omega = \pm 1 \}$
\bal \label{basis}
\cno =& \frac{1}{\sqrt{2}} \bigg(
c_{\bfk \uparrow}
+ \Omega \; \psi(\bfk) \; c_{\bfk \downarrow}  \bigg).
\end{align}
where we choose the convention that the basis $\{ \sigma = \uparrow,\downarrow \}$ is quantized in the $\hat{z}$ direction. We write the momentum-dependent amplitude $\psi$ as $\zeta(\bfk) + i\,\chi(\bfk)$, with $\zeta(\bfk)$ and $\,\chi(\bfk)$ real. The physical interpretations of $\zeta$ and $\chi$ are clear - a spinor in the x-y plane may be expressed as 
\bal \label{spin xy}
\frac{1}{\sqrt{2}}
\begin{pmatrix}
1 \\ \langle \sigma_x \rangle + i\, \langle \sigma_y \rangle
\end{pmatrix}.
\end{align}
The normalization $|\psi|^2 =1$ is just the condition that $\langle \sigma_x \rangle^2 + \langle \sigma_y \rangle^2=1$.\\

In assuming that the spin orientations of all electrons are coplanar, we are anticipating an exchange interaction that favors parallel alignment of spins. An out-of-plane degree of freedom in spin space will tend to disrupt the parallel alignment of spins and is energetically suppressed.\\ 

In the coplanar basis characterized by $\psi$, the kinetic hopping term has the form
\bal \label{ht}
H_t = -2 \, t \sum_{\bfk,\om} (\co k_x + \co k_y) \; \cdo \cno 
\end{align}
just like in the $\sigma$ basis. Before transforming to the $\Omega$ basis, the Coulomb interaction has terms quartic in creation and annihilation operators in the $\sigma$ basis:
\bal
H_U = &\frac{U}{2N} \,  \sum_{\substack{\bfk,\bfk',\bfq,\boldsymbol{\delta} \\ \sigma}}  c^{\dagger}_{\bfk + \bfq,\sigma} c^{\dagger}_{\bfk' - \bfq,-\sigma} c_{\bfk',-\sigma} c_{\bfk,\sigma}  \notag \\
H_J = &\frac{J}{2N} \,  \sum_{\substack{\bfk,\bfk',\bfq,\boldsymbol{\delta} \\ \sigma, \sigma'}} e^{i(\bfk - \bfk' + \bfq)\cdot \boldsymbol{\delta}} c^{\dagger}_{\bfk + \bfq,\sigma} c^{\dagger}_{\bfk' - \bfq,\sigma'} c_{\bfk',\sigma'} c_{\bfk,\sigma} \notag \\
H_{J'} = &\frac{J'}{2N} \,  \sum_{\substack{\bfk,\bfk',\bfq,\boldsymbol{\delta} \\ \sigma}} e^{i(\bfk + \bfk')\cdot \boldsymbol{\delta}} c^{\dagger}_{\bfk + \bfq,\sigma} c^{\dagger}_{\bfk' - \bfq,-\sigma} c_{\bfk',-\sigma} c_{\bfk,\sigma} \notag \\
H_V = &\frac{V}{2N} \,  \sum_{\substack{\bfk,\bfk',\bfq,\boldsymbol{\delta} \\ \sigma, \sigma'}} e^{i\bfq\cdot \boldsymbol{\delta}} c^{\dagger}_{\bfk + \bfq,\sigma} c^{\dagger}_{\bfk' - \bfq,\sigma'} c_{\bfk',\sigma'} c_{\bfk,\sigma} \notag \\
H_{\Delta t} = &\frac{\Delta t}{2N} \,  \sum_{\substack{\bfk,\bfk',\bfq,\boldsymbol{\delta} \\ \sigma}} \big( e^{i\bfk\cdot \boldsymbol{\delta}} + e^{i\bfk'\cdot \boldsymbol{\delta}} + e^{i(\bfk-\bfq)\cdot \boldsymbol{\delta}} + e^{i(\bfk'-\bfq)\cdot \boldsymbol{\delta}} \big)  \notag \\ & \times c^{\dagger}_{\bfk + \bfq,\sigma} c^{\dagger}_{\bfk' - \bfq,-\sigma} c_{_{\scriptstyle{\bfk',-\sigma}}} c_{\bfk,\sigma} .
\end{align}
We convert this to the $\Omega$ basis and decouple in the $\langle \cdo \cno \rangle$ direct channel. Define mean fields
\bal \label{nk}
n_{\bfk} =& n_{\bfk+} + n_{\bfk-} = \langle c^{\dagger}_{\bfk+} c_{\bfk+} \rangle + \langle c^{\dagger}_{\bfk-} c_{\bfk-}  \rangle \; \text{and}\\
\Delta_{\bfk} =& n_{\bfk+} - n_{\bfk-} = \langle c^{\dagger}_{\bfk+} c_{\bfk+} \rangle - \langle c^{\dagger}_{\bfk-} c_{\bfk-} \rangle. \label{delt}
\end{align}

In this paper we consider the set of ground states satisfying the symmetry constraints
\bal \label{no magn mome}
\sum_{\bfk} \psi(\bfk) \Delta_{\bfk} =& \; 0 \;\;\;\; \text{and} \\
\sum_{\bfk, \boldsymbol{\delta}} e^{i\,\bfk \cdot \boldsymbol{\delta}} \psi(\bfk) \Delta_{\bfk} =& \;0. \label{2nd symm}
\end{align}
We will call them `Class $J$ ground states'. We deduce from Eqs. (\ref{basis}) and (\ref{delt}) that Eq. (\ref{no magn mome}) sums over the spins of all the polarized electrons, hence these ground states have \emph{no net magnetic moment} - ferromagnetism does not belong in this class. Eq. (\ref{2nd symm}) ensures that the momentum distribution of the `polarized spin' $\psi\,\Delta$  does not have the symmetry of the underlying lattice; in two dimensions this symmetry is the point group $\text{C}_{n\nu}$ for a lattice with $n$-fold symmetry. From the choice of mean fields in Eqs. (\ref{nk}) and (\ref{delt}), we deduce that electron crystal momentum $\bfk$ is a good quantum number in Class $J$, hence antiferromagnetic and spin-density-wave states are excluded. \\

We prove in Sec. VI-C that these ground states break parity. For the special case that $\zeta(\bfk)\,\chi(\bfk)$ are invariant under $\bfk \rightarrow -\bfk$, we also show that they preserve time-reversal symmetry.  \\

There are many variations of ground states that fall into this class. We consider the following cases:\\ 




(i) $\Delta$ breaks the lattice symmetry but $\psi$ does not. The simplest example is when $\psi$ is independent of momentum - the spin orientation of all electrons point along the same axis - and there is a relative displacement of spin up and down Fermi surfaces with respect to each other. This is the spin-split state shown in Fig (1-a). \\

(ii) $\psi$ breaks the lattice symmetry but $\Delta$ does not. The simplest example is an expansion of one Fermi surface with respect to the other in a basis that rotates spin by $2\pi$ rad as we go around the Fermi contour, i.e. winding number of $1$. These include eigenstates of helicity (spin parallel or antiparallel to momentum) and eigenstates of the Rashba and Dresselhaus spin-orbit interactions. The Rashba state is shown in Fig (1-b).\\

(iii) It is also possible that $\psi$ and $\Delta$ both break the lattice symmetry. These exotic ground states are shown to be energetically unfavorable in Sec. VI-B. \\ 

These constraint will have important implications on which Hubbard parameters are relevant to the energetics of polarization. In particular, we will find that the nearest-neighbor exchange $J$ is the greatest driving force for symmetry breaking, hence the name Class $J$.\\

We proceed to reduce the Hamiltonian in Eq. (\ref{hami}). We introduce matrices
\bal \label{pi}
\Pi(\bfk') = &n_{\bfk'}\bpm 1 & 0 \\ 0 & 1 \epm, \\
\Upsilon(\bfk, \bfk') = & -\dk \bpm \big( \zeta \, \zeta' + \chi\,\chi' \big) & \; i \big( \chi \, \zeta' - \zeta\,\chi' \big)  \\[0.6em] -i \big( \chi \, \zeta' - \zeta\,\chi' \big)  & \;-\big( \zeta \, \zeta' + \chi \,\chi' \big)  \epm. \label{upsi}
\end{align}
To simplify notation, we have written $\zeta_{\bfk}$ as $\zeta$ and $\zeta(\bfk')$ as $\zeta'$ in Eq. (\ref{upsi}),
and similarly for $\chi$. For the rest of this paper, we use the convention that $\Pi = \Pi(\bfk')$ and $\Upsilon = \Upsilon(\bfk,\bfk')$ as defined in Eqs. (\ref{pi}) and (\ref{upsi}). The diagonal and off-diagonal elements of $\Upsilon$ are proportional to $\langle \boldsymbol{\sigma} \rangle \cdot \langle \boldsymbol{\sigma}' \rangle$ and $\langle \boldsymbol{\sigma} \rangle \times \langle \boldsymbol{\sigma}' \rangle$ respectively. \\

There are two values of momentum transfer $\bfq$ that contribute to the mean field: $\bfq = \boldsymbol{0}$ and $ \bfq = \bfk' - \bfk $. For example, consider the on-site repulsion $U$. We may organize 
\bal
H_U = H_U^{\scriptscriptstyle{(\bfq = \boldsymbol{0})}} + H_U^{\scriptscriptstyle{(\bfq = \bfk'-\bfk)}}.
\end{align} 
After the mean-field decoupling,
\bal \label{hu}
H_U^{\scriptscriptstyle{(\bfq = \boldsymbol{0})}} = &\frac{U}{2N} \sum_{\bfk,\bfk',\boldsymbol{\delta}} \begin{pmatrix} \cdp & \cdm \end{pmatrix} \bigg( \Pi \bigg) \bpm \cnp \\ \cnm \epm  \\
H_U^{\scriptscriptstyle{(\bfq = \bfk' - \bfk)}} = &\frac{U}{2N} \sum_{\bfk,\bfk',\boldsymbol{\delta}} \begin{pmatrix} \cdp & \cdm \end{pmatrix} \bigg( \Upsilon \bigg) \bpm \cnp \\ \cnm \epm. \label{hu 2}
\end{align}
We have neglected constants proportional to $\langle \cda c \rangle \langle \cda c \rangle$ that will affect the expectation value of the ground state energy but not the \text{sin}gle-particle energy spectrum. We carry out the integral over $\bfk'$. Defining $n$ as the total number of electrons, $\sum_{\bfk'} \Pi = n \; \delta_{\eta \eta'}$ which is a constant matrix independent of polarization. Eq. (\ref{no magn mome}) leads to $\sum_{\bfk'} \Upsilon = 0$. We conclude that $H_U$ does not contribute to Class $J$-type polarization. For the same reasons, we may discard $H_J^{\scriptscriptstyle{(\bfq = \bfk' - \bfk)}}$ and $H_V^{\scriptscriptstyle{(\bfq = \boldsymbol{0})}}$. \\

We also consider
\bal
&H_{\Delta t}^{\scriptscriptstyle{(\bfq = \bfk' - \bfk)}} \notag \\
= &\frac{\Delta t}{2N} \sum_{\bfk,\bfk',\boldsymbol{\delta}} \big( e^{i\bfk \cdot \boldsymbol{\delta}}  + e^{i\bfk' \cdot \boldsymbol{\delta}} \big) \begin{pmatrix} \cdp & \cdm \end{pmatrix} \bigg( 2\Upsilon \bigg) \bpm \cnp \\ \cnm \epm.
\end{align}
The sum over $\bfk'$ vanishes because of both symmetry constraints in Eq. (\ref{no magn mome}) and (\ref{2nd symm}).\\

The following terms are left:
\bal \label{hj}
&H_J^{\scriptscriptstyle{(\bfq = \boldsymbol{0})}} \notag \\
= &\frac{J}{2N} \sum_{\bfk,\bfk',\boldsymbol{\delta}} e^{i(\bfk - \bfk')\cdot \boldsymbol{\delta}}  \begin{pmatrix} \cdp & \cdm \end{pmatrix} \bigg( 2\Pi \bigg) \bpm \cnp \\ \cnm \epm
\end{align}
\bal \label{hj'}
&H_{J'} \notag \\
= &\frac{J'}{2N} \,  \sum_{\bfk,\bfk',\boldsymbol{\delta}} e^{i(\bfk + \bfk')\cdot \boldsymbol{\delta}} \begin{pmatrix} \cdp & \cdm \end{pmatrix} \bigg( \Upsilon + \Pi \bigg) \bpm \cnp \\ \cnm \epm
\end{align}
\bal \label{hv}
&H_V^{\scriptscriptstyle{(\bfq = \bfk' - \bfk)}} \notag \\
= &\frac{V}{2N} \sum_{\bfk,\bfk',\boldsymbol{\delta}} e^{i(\bfk - \bfk')\cdot \boldsymbol{\delta}} \begin{pmatrix} \cdp & \cdm \end{pmatrix} \bigg( \Upsilon - \Pi \bigg) \bpm \cnp \\ \cnm \epm
\end{align}
\bal
&H_{\Delta t}^{\scriptscriptstyle{(\bfq = \boldsymbol{0})}} \notag \\
= &\frac{\Delta t}{2N} \sum_{\bfk,\bfk',\boldsymbol{\delta}} \big( e^{i\bfk \cdot \boldsymbol{\delta}}  + e^{i\bfk' \cdot \boldsymbol{\delta}} \big) \begin{pmatrix} \cdp & \cdm \end{pmatrix} \bigg( 2\Pi \bigg) \bpm \cnp \\ \cnm \epm.
\end{align}
With the exception of $H_{\Delta t}^{\scriptscriptstyle{(\bfq = \boldsymbol{0})}}$, the other interactions are particle-hole symmetric - we will show that their effects are extremized at half-filling.\\

We briefly describe the roles of the various Hubbard parameters:\\

(i) The expectation value of Eq. (\ref{hj}) is
\bal \label{expe hj}
\langle H_J^{\scriptscriptstyle{(\bfq = \boldsymbol{0})}} \rangle = \frac{J}{2N} \,  \sum_{\bfk,\bfk',\boldsymbol{\delta}} e^{i(\bfk - \bfk')\cdot \boldsymbol{\delta}} \; n_{\bfk} \;n_{\bfk'}.
\end{align}
The amplitude $\sum_{\boldsymbol{\delta}}\;e^{i(\bfk - \bfk')\cdot \boldsymbol{\delta}}$ is negative when $k_i \approx 0$ and $k'_i \approx \pm \pi$ or vice versa, hence $J$ favors a separation of electrons in momentum space, i.e. polarization . This was first pointed out by one of us \cite{hirschfm} in connection with ferromagnetism. Since Eq. (\ref{expe hj}) is independent of the choice of basis, we find that $J$ promotes polarization in any arbitrary basis; this is true even for ground states that do not satisfy the symmetry constraints in Eq. (\ref{no magn mome}) and (\ref{2nd symm}). Class $J$ ground states are special because they reduce their energy \emph{most effectively} through nearest-neighbor exchange $J$ when they break symmetry. \\

(ii) The expectation value of Eq. (\ref{hv}) is
\bal \label{expe hv}
&\langle H_V^{\scriptscriptstyle{(\bfq = \bfk' - \bfk)}} \rangle \notag \\
= &-\frac{V}{4N} \sum_{\bfk,\bfk',\boldsymbol{\delta}} e^{i(\bfk - \bfk')\cdot \boldsymbol{\delta}} \bigg( \Delta_{\bfk} \Delta_{\bfk'} \langle \boldsymbol{\sigma} \rangle \cdot \langle \boldsymbol{\sigma}' \rangle + n_{\bfk} \;n_{\bfk'} \bigg).
\end{align}
The second term has the opposite effect of $J$ and tends to suppress polarization. The first term causes an energy splitting between the $\Omega = \pm 1$ bands and tends to lower the energy if polarized electrons with $\bfk \approx \bfk'$ have parallel spin. This is analogous to the exchange energy $J$ that favors parallel spins for $localized$ electrons; this was proposed by Heisenberg as a mechanism to explain ferromagnetism \cite{Hei}. The difference is that in the momentum representation, $J$ influences the direct energy while $V$ mixes both direct and exchange energy.\\

While $J$ unambiguously assists polarization in any basis, $V$ affects both direct and exchange energies in a manner that is self-cancelling; its net effect on polarization depends sensitively on (i) the choice of ground state and (ii) band filling. Hence, $V$ plays the crucial role of deciding which ground state is energetically favored for a given band filling.\\

(iii) We show in Sec. III-E that the main effects of $J'$ and $\Delta t$ are to cancel $V$ and $t$ respectively.\\
 
We now analyze the two simplest examples of Class $J$ with this reduced Hamiltonian - they are the Rashba-type and the spin-split states. In Sec. VI-B, we demonstrate that they are also the most stable in this class. \\

\section{Case study of Rashba-type ground state}

We examine a set of ground states with spin structures that diagonalize Hamiltonians of the form $H \propto \sigma^x \, \si k_a \pm \sigma^y \, \si k_b$. For specificity, we first consider the Rashba spin-orbit basis. For completeness of analysis, we add a Rashba single-particle interaction $H_R$
to the Hamiltonian in Eq. (\ref{hami}); this originates from the Dirac spin-orbit interaction (Eq. (\ref{rash free hami})) with an electric field perpendicular to the plane
\bal
H = H_t + H_R + H_U + H_J + H_{J'} + H_V + H_{\Delta t}.
\end{align}
With only nearest-neighbor hopping, the tight-binding version of the single-particle Rashba Hamiltonian (Eq. (\ref{rash free hami})) is \cite{mireles}
\bal \label{rash hami}
H_R = -\R \; \sum_{\bfk \sigma \sigma'} c_{\bfk \sigma}^{\dagger} \bigg( \si k_{x}\big( \sigma^y \big)_{_{\sigma \sigma'}} - \si k_y \big( \sigma^x \big)_{_{\sigma \sigma'}} \bigg) c_{\bfk \sigma'} \;\;
\end{align}
where spin indices $\sigma$ are chosen to be diagonal in the $\hat{z}$ direction; Eq. (\ref{rash hami}) is formally derived in App. A. Diagonalizing this Hamiltonian in momentum space, we obtain
\bal
H_R = \R \sum_{\bfk,\eta} \eta \sqrt{\si^2k_x +\si^2k_y} \; \cdg \cnd. 
\end{align}
We distinguish between indices $\sigma$ and $\eta$; the latter denotes the Rashba polarization and takes on values $\pm 1$. The eigenvectors are 
\bal \label{ck}
\cnd =& \frac{1}{\sqrt{2}} \bigg(
c_{\bfk \uparrow}
+ \eta \frac{f_k}{|f_k|} c_{\bfk \downarrow}  \bigg)
; \\ \label{fk}
f_k =& \; \si k_y + i \,\si k_x. 
\end{align}
 
We introduce the generalized dot and cross products
\beqn
\tki = \frac{sin \, k_i}{\sqrt{\sum_j sin^2 \, k_j}},\\
\tilde{k} \cdot \tilde{k'} = \sum_{ij} \tki \tkj \delta_{ij},\\
\tilde{k} \times \tilde{k'} = \sum_{ij} \tki \tkj \epsilon_{ij}.
\eeqn
Note that $\epsilon_{ij}$ is the Levi-Cevita symbol. The matrices introduced in Eq. (\ref{pi}) and (\ref{upsi}) take the form
\bal
\Pi = &n_{\bfk'}\bpm 1 & 0 \\ 0 & 1 \epm \\
\Upsilon = & \dk \bpm -\big( \tk \cdot \tkp \big) & \; -i \big( \tk \times \tkp \big)  \\[0.6em] i \big( \tk \times \tkp \big)  & \;\big( \tk \cdot \tkp\big)  \epm.
\end{align}

We may easily generalize the above results to include the tight-binding generalizations of the Dresselhaus and helicity spin-orbit bases. The Dresselhaus basis follows from changing the spin orientation $f_k \rightarrow \si k_x + i\, \si k_y$ in Eqs. (\ref{ck}) and (\ref{fk}); the helicity basis changes $f_k \rightarrow \si k_x - i\, \si k_y$. Without any single-particle spin-orbit coupling, all three ground states are degenerate because electron-electron interactions cannot distinguish between their spin structures. Other than the spin structure encoded in $f_k$, all the results that follow in the rest of Sec. III are applicable to the Dresselhaus and helicity states if $\alpha_R=0$. \\

We first analyze a further reduced Hamiltonian with  only $H_t,H_R,H_J^{\scriptscriptstyle{(\bfq = \boldsymbol{0})}} $ and $H_V^{\scriptscriptstyle{(\bfq = \bfk' - \bfk)}}$ in one and two dimensions; we defer a discussion of the effects of $H_{\Delta t}^{\scriptscriptstyle{(\bfq = \boldsymbol{0})}}$ and $H_{J'}$ to Sec. III-E.\\

\subsection{One Dimension}

The reduced Hamiltonian is exactly solved in one dimension. Many of the qualitative conclusions generalize to two dimensions.\\

The generalized dot product reduces to $\tilde{k} \cdot \tilde{k'} = \, \text{sgn}(k) \, \text{sgn}(k')$ and the cross product vanishes. The interaction part of the Hamiltonian simplifies to 

\bal
H_{V,J} =& \frac{1}{N} \,  \sum_{\bfk,\bfk',\eta} \text{cos}(k-k') c^{\dag}_{k\eta} c_{k\eta} \notag \\
&\times \bigg( (2J-V)n_{k'} - \eta V \Delta_{k'} \, \text{sgn}(k) \, \text{sgn}(k') \bigg) 
\end{align}

We parametrize the polarization by a dimensionless quantity $\epsilon$: $n_{k \pm} = 1$ for $k \in \{ -k_{\scriptscriptstyle{F}} \pm \epsilon, +k_{\scriptscriptstyle{F}} \pm \epsilon \}$. The \text{sin}gle-particle energy is

\bal \label{disp rash 1d}
E_\eta (k) = &-2\left( t - \frac{1}{\pi}(2J-V) \,\text{sin} \,k_{\scriptscriptstyle{F}} \,\text{cos} \, \epsilon \, \right) \text{cos} \,k\,  \notag \\
&+\eta \left( \R + \frac{2}{\pi} V \,\text{sin}\,k_{\scriptscriptstyle{F}} \, \text{sin}\, \epsilon \right) \text{sin} \, |k|.
\end{align}

The ground state expectation value of the full Hamiltonian is

\bal \label{ener rash 1d}
\frac{\langle H \rangle}{N} = &-\frac{2}{\pi} \;\text{sin}\, k_{\scriptscriptstyle{F}} \bigg( \R \, \text{sin}\, \epsilon + 2\,t\,\text{cos}\, \epsilon \bigg) \notag \\
&+ \frac{2}{\pi^2}\;\text{sin}^2\, k_{\scriptscriptstyle{F}} \bigg(2J\, \text{cos}^2 \, \epsilon - V \bigg). 
\end{align}

From Eqs. (\ref{disp rash 1d}) and (\ref{ener rash 1d}), we see explicitly that $J$ promotes polarization by increa\text{sin}g the bandwidth of the \text{sin}gle-particle energy spectrum. On polarization, $V$ simultaneously reduces the bandwidth and increases the energy splitting of the Rashba bands; the net effect is that the $V$ term in the ground state energy is independent of $\epsilon$. \\

Minimizing $\langle H \rangle$ with respect to $\epsilon$, the ground state polarization $\epsilon^*$ is determined by

\bal
\R \, \text{\text{cos}} \, \epsilon^* + \frac{2}{\pi} J \, \text{\text{sin}}\, k_{\scriptscriptstyle{F}} \, \text{\text{sin}} \, (2\epsilon^*) = 2t\, \text{\text{sin}} \, \epsilon^*.
\end{align}  
This same equation can be obtained by enforcing the self-consistency condition
\bal \label{self cons 1d}
E_- (k_{\scriptscriptstyle{F}}+\epsilon^*) = E_+ (k_{\scriptscriptstyle{F}} - \epsilon^*).
\end{align}
In Sec. III-B, we prove that they are equivalent statements by minimization with Lagrange multipliers.\\

If $\R=0$, the polarization is determined by

\bal
\text{cos} \, \epsilon^* = \frac{\pi}{2} \frac{t}{J\, \text{sin}\, k_{\scriptscriptstyle{F}}}
\end{align}
and there is no nonzero solution for $J < J_c$ with 
\bal \label{jc rash 1d}
J_c = \frac{\pi}{2} \frac{t}{ \text{sin}\, k_{\scriptscriptstyle{F}}}.
\end{align}
At $J>J_c$ there is a symmetry breaking: $\epsilon$ may take on values $\pm \epsilon^*$. Note that the effect of $J$ is maximized at half-filling; this is consistent with the particle-hole symmetry of the nearest-neighbor exchange interaction. \\

Up to an arbitrary spin rotation, the Rashba state is equivalent to the spin-split state in one dimension \cite{hirschss}; we show this in Sec. IV-A.\\

\subsection{Two Dimensions}

\begin{figure}[t]
		\subfigure[]{\resizebox{8cm}{!}{\includegraphics[width=7cm, height=7cm]{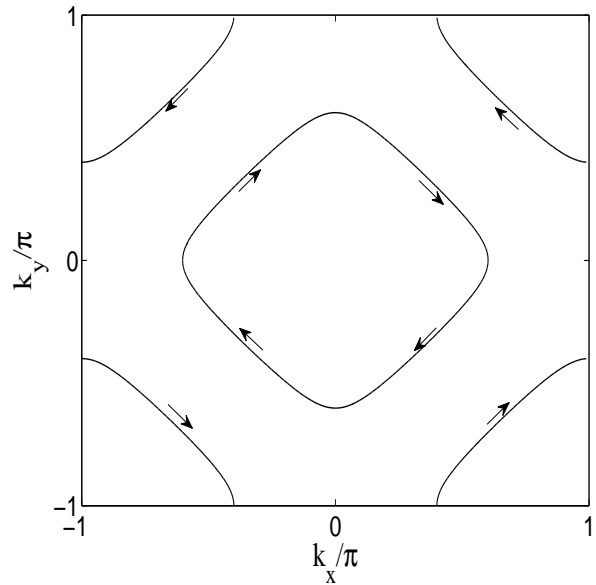}}}
		\subfigure[]{\resizebox{8cm}{!}{\includegraphics[width=7cm, height=7cm]{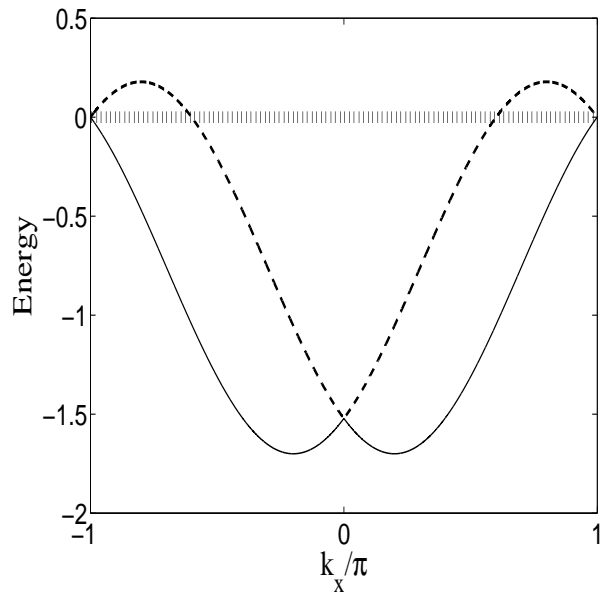}}}
		\caption{(a) \, Fermi energy contours of the two Rashba bands. Arrows refer to spin directions. (b) \, Dispersion at fixed $k_y=0$.  Parameters: $t=1, V=2, J=3, \R=0$. This band is half-filled. Symmetry breaking occurs for $J>J_c = 2.5$.}
\end{figure}

The off-diagonal terms in $H_{V,J}$ are proportional to the integral
\bal
\sum_{\boldsymbol{\delta},\bfk'} e^{i(\bfk - \bfk')\cdot \boldsymbol{\delta}} \; (\tk \times \tkp) \; \dk,
\end{align}
which vanishes if the splitting satisfies $\Dk = \Delta_{-\boldsymbol{k}}$ and $\Delta_{k_x,k_y}=\Delta_{\pm k_y,\pm k_x}$. The latter condition reflects the symmetry of the underlying square lattice.\\

We define the geometric factors
\bal \label{alph}
\alpha =& \frac{1}{N} \sum_{\bfk} (\text{cos}\,k_x + \text{cos}\, k_y) \; n_{\bfk}, \\
\beta =& \frac{1}{N} \sum_{\bfk} \sqrt{\text{sin}^2\,k_x + \text{sin}^2 \, k_y} \; \Delta_{\bfk}. \label{beta}
\end{align}
Up to a negative constant $-2t$, $\alpha$ is the sum of the kinetic energies of all the electrons; $\beta$ can be interpreted as the order parameter and is defined to be negative for positive $\R$. We also define $\alpha_o$ as the value of $\alpha$ when there is zero polarization; $\beta_o = 0$.\\

The single-particle energy spectrum is then
\bal \label{disp rash}
E_{\eta}(k) = &-2\bigg( t - \frac{1}{4} (2J-V) \alpha \bigg) (\text{cos}\,k_x + \text{cos}\, k_y) \notag \\
&+ \eta \bigg( \R - \frac{1}{2} V \beta \bigg) \sqrt{\text{sin}^2\,k_x + \text{sin}^2 \, k_y}
\end{align}
We observe that $V$ results in an effective momentum-dependent Zeeman field that is maximized at momenta ($\pm \pi/2, \pm \pi/2$) while $\beta$ is the overlap of the polarization $\Delta_{\bfk}$ with this Zeeman field. Consequently, $V$ reduces the exchange energy most effectively at half-filling, where any polarization takes maximal advantage of the Zeeman field.  \\

\begin{figure}[b]
 \begin{center}
 \subfigure[]{\resizebox{8cm}{!}{\includegraphics[width=7cm, height=7cm]{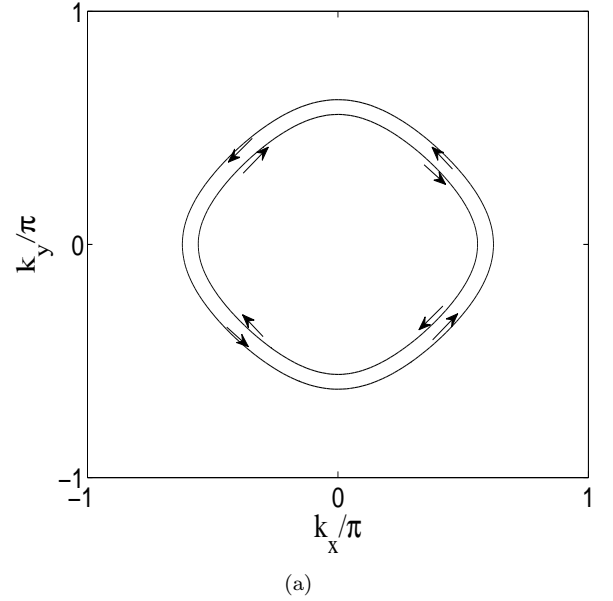}}}
		\subfigure[]{\resizebox{8cm}{!}{\includegraphics[width=7cm, height=7cm]{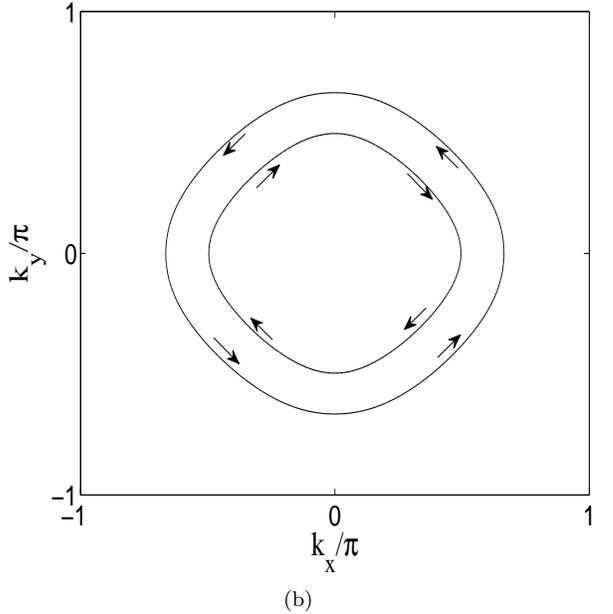}}}
		\end{center}
		\caption{(a) \, Fermi energy contours of the two Rashba bands for $J=0$ (b) \, Fermi energy contours of the two Rashba bands for $J=2$. Filling $=0.5$. Parameters: $t=1, V=0, \, \R=0.2$. }
\end{figure}

The ground state energy is
\bal\label{ener rash}
\frac{\langle H \rangle}{N} = -2t\alpha + \R \beta  + \frac{1}{2} J \alpha^2 - \frac{1}{4} V (\alpha^2 + \beta^2). 
\end{align}
$\alpha$ ($\beta$) increases (decreases) as the system polarizes; by inspection of Eq. (\ref{ener rash}), we confirm that $J$ and $\R$ favor polarization. Whether $V$ assists polarization or not is not clear from Eq. (\ref{ener rash}) alone. In one dimension, $\alpha^2+\beta^2$ is a constant that is independent of polarization (c.f. Eq. (\ref{ener rash 1d})), hence $V$ will not tip the balance in either direction. In Sec. III-C and V, we investigate the behavior of $\alpha^2+\beta^2$ in two dimensions.\\

We wish to minimize the ground state energy with the constraint of fixed particle number $n$. Hence we define
\beq \label{defi G}
G = \; \frac{\langle H \rangle}{N} - \mu \left( \sum_{\bfk} n_{\bfk} -n \right)
\eeq
and impose the condition
\beq
\frac{\partial G}{\partial n_{\bfk \eta}} = \; 0.  \label{mini G}
\eeq
Since we are making an infinitesimal variation to the Fermi surfaces of both Rashba bands, Eq. (\ref{mini G}) only applies to states at the two Fermi surfaces. From Eq. (\ref{mini G}),
\bal \label{self cons 2d}
\big( E_\eta (\bfk) \big)_{FS, \eta} = \mu
\end{align}    
or
\bal \label{ener mini cond}
\big( E_+ (\bfk) \big)_{FS+} = \big( E_- (\bfk') \big)_{FS-}
\end{align}
which is a generalization of Eq. (\ref{self cons 1d}). \\

We define $a$ and $b$:
\bal \label{a b}
E_{\bfk \eta} = - a (\text{cos}\,k_x + \text{cos}\,k_y) + \eta \; b \sqrt{\text{sin}^2k_x +\text{sin}^2k_y}.
\end{align}
$b$ is the energy splitting and $a$ is a quarter of the bandwidth; for convenience, we call $a$ the bandwidth. \\

Given parameters $t,J,V,\R$ and a filling $n$, Eqs. (\ref{nk}),(\ref{delt}),(\ref{alph}),(\ref{beta}), (\ref{disp rash}) and (\ref{self cons 2d}) may be solved numerically. A few representative results are presented: (i) in Fig. (2-a) and (2-b), a symmetry-breaking solution exists for $\R=0$ and $J>J_c$  (ii) in Fig. (3-a) and (3-b), a comparison is made between two Rashba-polarized systems with equal $\R$ but different $J$.

\subsection{A useful analytic approximation}

We are motivated by the simplicity of the equations in one dimension to look for a suitable approximation of the two-dimensional equations; this approximation permits us to extract an analytic formula that relates $J,\R$ and $t$. We numerically explore the limits of validity of this approximation. Where the approximation is not valid, we believe our conclusions are at least qualitatively correct.\\

In one dimension, the Rashba bands can polarize in a \text{sin}gle direction. Hence, we minimize $\langle H \rangle$ with respect to a \text{sin}gle polarization parameter $\epsilon$ to obtain the unique ground state with polarization $\epsilon^*$. In addition,
\bal \label{alph 1d}
\alpha_{_{1D}} =& \frac{4}{\pi} \text{sin} \, k_{\scriptscriptstyle{F}} \, \text{cos}\, \epsilon \\
\beta_{_{1D}} =& - \frac{4}{\pi} \text{sin} \, k_{\scriptscriptstyle{F}} \, \text{sin}\, \epsilon \label{beta 1d}
\end{align}
and the combination $\alpha_{\scriptscriptstyle{1D}}^2 + \beta_{\scriptscriptstyle{1D}}^2$ is independent of polarization parameter $\epsilon^*$ and only dependent on the the number of electrons; this is a consequence of number conservation. \\

In two dimensions, the Rashba bands can polarize anisotropically, hence we have to minimize $\langle H \rangle$ with respect to all the occupation numbers on the Rashba Fermi surfaces (see Eq. (\ref{mini G})). This minimization process is equivalent to minimizing $\langle H \rangle$ with respect to mean fields $\alpha$ and $\beta$; this claim is justified by the Hohenberg-Kohn theorem \cite{hoh} in App. B. After minimizing Eq. (\ref{ener rash}), we obtain
\bal \label{mini rash}
2\,t = \bigg( J - \frac{V}{2}\bigg) \alpha +  \left( \R - \frac{V}{2} \beta \right) \frac{\delta \, \beta}{\delta \, \alpha}.
\end{align}
$\delta \, \beta/ \delta \, \alpha$ is fixed by the constraint of number conservation and is generally a complicated function of $\alpha$ and $\beta$. The constancy of $\alpha_{\scriptscriptstyle{1D}}^2 + \beta_{\scriptscriptstyle{1D}}^2$ in one dimension suggests that we look for a similar condition in two dimensions. In the limit that the polarization is small relative to the filling, numerical results confirm that 
\bal\label{anal appr}
\frac{\delta \, \beta}{\delta \, \alpha} \approx - \frac{\alpha}{\beta}.
\end{align}
Eq. (\ref{anal appr}) is a good approximation even when extended to systems that are medium-polarized ($<50\%$ of the total filling); this is illustrated in Fig. (4). In Sec. V-A, we provide a geometric explanation of why this approximation is especially good near quarter-filling.\\

\begin{figure}[b]
		\resizebox{8.5cm}{!}{\includegraphics[width=8cm]{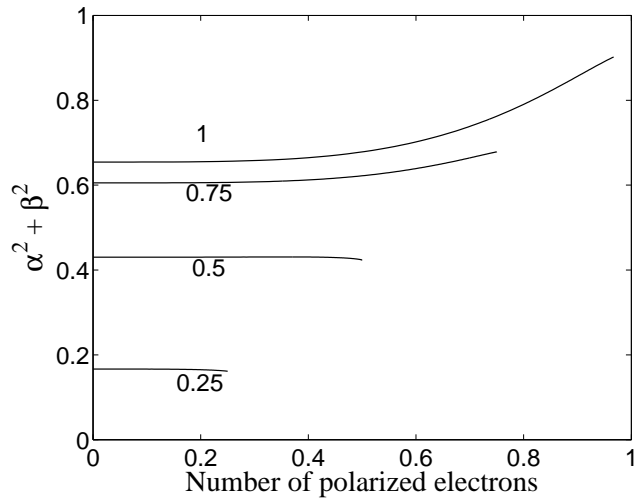}}
	\caption{ $\alpha^2 + \beta^2$ vs number of polarized electrons. Parameters are $t=1, J=1, V=1$ and $\R$ is varied to change the extent of polarization. Numbers $0.25,0.5,0.75$ and $1$ on the graph refer to the total number of electrons. $\alpha^2 + \beta^2$ can be approximated as a constant until the number of polarized electrons exceeds half the total number of electrons. }
	\label{ab vs O}
\end{figure}

Employing Eq. (\ref{anal appr}), we cast Eq. (\ref{mini rash}) in the useful form
\bal \label{master}
J = \frac{1}{\beta} \R + \frac{2}{\alpha} t.
\end{align}
We note that $\beta$ is nonzero for any finite $\R$ - a single-particle Rashba interaction always polarizes the system. We may minimize Eq. (\ref{ener rash}) with $\R$ strictly zero to obtain the critical $J$ for symmetry breaking
\bal \label{jc rash}
J_c =  \frac{2}{\alpha_o} t 
\end{align}
which is consistent with Eq. (\ref{jc rash 1d}) in one dimension. Beyond the approximation in Eq. (\ref{anal appr}), we examine a correction to Eq. (\ref{jc rash}) in Sec. V-B and find that it is small for moderate values of $V$. For a half-filled band in the unpolarized state, electrons occupy all momenta with positive $\co k_x\, +\, \co k_y $, hence symmetry breaking is most likely to occur at half filling where $\alpha_o$ is maximized. We derive in App. C the following equation 
\bal
\frac{J_c}{t} = &\;\frac{\pi}{k_{\scriptscriptstyle{F}} \; J_1(k_{\scriptscriptstyle{F}})} \notag \\
= &\;(2\pi) \, \bigg( k_{\scriptscriptstyle{F}}^2 - \frac{k_{\scriptscriptstyle{F}}^4}{8} + \frac{k_{\scriptscriptstyle{F}}^6}{768} - \frac{k_{\scriptscriptstyle{F}}^8}{9216} + \dots \bigg)^{-1}
\end{align}
that is valid for small filling; $J_1(k_{\scriptscriptstyle{F}})$ is a Bessel function of the first kind, of integral order $1$. In this limit we see explicitly that $J_c$ is lowered when filling increases. For $\R=0$ and $J<J_c$, we may express the bandwidth defined in Eq. (\ref{disp rash}) and (\ref{a b}) as
\bal
a = \alpha_o \; \bigg( \frac{2t}{\alpha_o} - J + \frac{V}{2} \bigg).
\end{align}
As $J$ is increased, symmetry breaking prevents the bandwidth from reaching the critical value of $\alpha_0 V/2$.\\

\begin{figure}[h]
		\resizebox{8.5cm}{!}{\includegraphics[width=8cm]{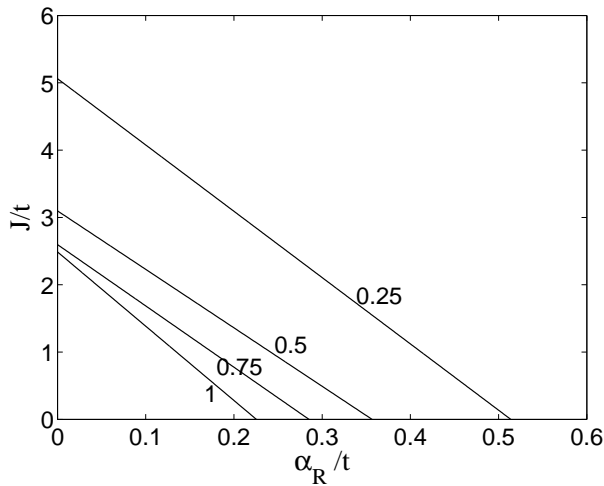}}
		\caption{Analytic approximation of $J/t$ vs $\R/t$. These are the values of $J,\R$ and $t$ consistent with a polarization of $0.1$ for fillings $0.25,0.5,0.75$ and $1$.}
	\label{j vs R}
\end{figure}

We observe that $\alpha$ and $\beta$ are just the band filling and polarization weighted by different geometric factors in their respective integrals. For fixed $\alpha$ and $\beta$, we may ask what are the range of $J/t$ and $\R/t$ that are consistent with this choice of polarization and filling. Eq. (\ref{master}) gives us a simple linear relation between $J/t$ and $\R/t$ that we plot in Fig. (5) for various fillings. A Rashba-polarized system that is conventionally explained by a nonzero $\R$ is indistinguishable from a system with a smaller (even zero) $\R$ but larger $J$ -  there is a trade-off between $\R$ and $J$ which is characterized by the slope $1/\beta$. Typically, $-\beta\, \sim \text{O}(\bar{p})$ with $\bar{p}$ the number of polarized electrons normalized to a maximum of $1$; this maximum occurs in the fully-polarized half-filled band. We explore the implications of Eq. (\ref{master}) for \emph{Au}(111) in Sec. III-D.\\

The analytic approximation in Eq. (\ref{anal appr}) is equivalent to the statement that the polarization is independent of $V$. We have shown that this approximation is only consistent if the resultant polarization is small relative to the band filling. We defer a discussion of the role of $V$ beyond this approximation to Sec. V-B.\\

\subsection{Comparison with \emph{Au}(111)}

We investigate the range of values of ${t,V,J,\R}$ that are compatible with the observed polarization of \emph{Au}(111) surface states, which has a hexagonal Brillouin zone. The small filling suggests we may approximate the Brillouin zone as square. \\

\emph{Au}(111) has a lattice constant of $2.89\mathrm{\AA}$. The Fermi momenta of the two Rashba bands are $0.153\A$ and $0.177\A$\cite{au1}. The energy dispersion is fitted to $E(k) = 1.82\text{eV} (k \pm 0.0338)^2$. In the low-$k$ limit, the tight-binding dispersion reduces to $E = -a(2-k^2/2) \pm b\,k$ from which we infer $a \approx 3.64\text{eV}$ and $b \approx 0.123\text{eV}$. $\alpha$ and $\beta$ are evaluated by numerically evaluating Eq. (\ref{alph}) and (\ref{beta}) assuming circular Fermi surfaces: $\alpha = 0.0707, \beta = -0.00244$.\\

We define $\bar{n}$ ($\bar{p}$) as the band filling (polarization) normalized to the range $[0,2] ([0,1])$. Given the above values of $\alpha$ and $\beta$, we numerically solve the self-consistent mean field equations to obtain $\bar{n} = 0.0363$ and $\bar{p}=0.00525$. This can be compared with the fractional filling and polarization obtained if we assume the experimental dispersion is embedded in a square lattice, i.e. $\bar{n}= \pi (0.153^2 + 0.177^2)/ (2\pi)^2 = 0.0364$ and $\bar{p} = \pi (0.177^2 -0.153^2) / (2\pi)^2 = 0.00526$. \\

Since the polarization is small compared to the filling, we may employ the approximation of Eq. (\ref{anal appr}). From Eq. (\ref{master}), $J/t = 28.3 - 410 \, \R/t$; this is plotted in Fig. (6).\\

\begin{figure}[h]
		\resizebox{8.5cm}{!}{\includegraphics[width=8cm]{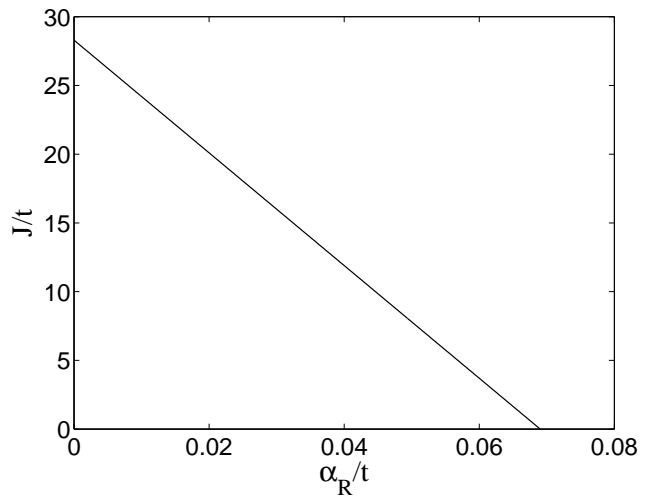}}
		\caption{$J/t$ vs $\R/t$ in the model of \emph{Au}(111).}
	\label{j vs R for Au}
\end{figure}

The conventional understanding of \emph{Au}(111) is equivalent to setting $J=V=0$, hence $\R/t = 0.069$. We ask if the experimental data is consistent with a smaller value of $\R$.\\

(i) From Eq. (\ref{a b}), the energy splitting is related to the Rashba constant and $\beta$ by: $\R + 0.00122\,V \approx 0.123 \,\text{eV}$. Reducing $\R$ requires that $V$ be unphysically large if the momentum-space polarization $\beta$ is small. As per our discussion in Sec. III-B, the effect of $V$ is maximized at half-filling; $Au(111)$'s band is near-empty, hence $V$ by itself is too weak to produce the required energy splitting. \\

(ii) The energy-minimizing condition in Eq. (\ref{master}) relates $J,\,\R,\,t,\,\alpha,\,\beta$: $J/t = 28.3 - 410 \, \R/t$. Reducing $\R$ while keeping $t$ fixed requires that $J$ be unphysically large. This is because $J$ most effectively polarizes a band at half-filling but the \emph{Au}(111) band is nearly empty. One option to avoid enlarging $J$ is to decrease both $\R$ and $t$ simultaneously. However, this option is costly in $V$ as per our discussion in (i).\\

We conclude that it is unlikely that $J$ and $V$ play a significant role in \emph{Au}(111). \\

\subsection{Effects of $\Delta t$ and $J'$}

The near-identical structures of Eq. (\ref{hj'}) and Eq. (\ref{hv}) (except for minus signs) suggest the oppo\text{sin}g roles played by $J'$ and $V$. One can show that the only effect of $J'$ is to cancel $V$. We may replace 
\bal \label{veff}
V \longrightarrow V_{\scriptscriptstyle{\text{eff}}} = V - J'
\end{align}   
in the preceding analysis. On polarization, $J'$  helps to increase the  bandwidth but reduces the energy splitting - its effects are self-cancelling.\\

The effect of $\Delta t$ on the energy spectrum is
\bal
E_{\eta}(\bfk) \longrightarrow E_{\eta}(\bfk) + 2\Delta t \bigg( \alpha + \bar{n} \big( \text{\text{cos}}\,k_x + \text{\text{cos}} \,k_y \big)\bigg).
\end{align}
$\Delta t$ promotes polarization in two ways:\\

(i) The hopping parameter $t$ is renormalized to $t_{\scriptscriptstyle{\text{eff}}} = t - \Delta t \, \bar{n}$. This reduces the kinetic energy \text{cos}t of polarization. $\Delta t$ is the only nearest-neighbor interaction that breaks particle-hole symmetry on a square lattice - this is reflected on the dependence of $t_{\scriptscriptstyle{\text{eff}}}$ on $\bar{n}$. \\

(ii) On polarization, $\alpha$ decreases and the energies of all electrons are reduced by the same amount, independent of $\bfk$.\\

The effect of $\Delta t$ on the ground state energy is
\bal
\frac{\langle H \rangle}{N} \longrightarrow \frac{\langle H \rangle}{N} + 2\Delta t \, \bar{n} \, \alpha.
\end{align}
The energy minimizing values of $\alpha$ and $\beta$ are related by 
\bal
J = \frac{1}{\beta} \R + \frac{2}{\alpha} \big(t -\Delta t \, \bar{n} \big)
\end{align} 
with the approximation of Eq. (\ref{anal appr}).

\section{Case study of the Spin-Split State}

In Ref. \cite{hirschss} a parity-breaking spin-split Fermi-surface instability was studied with a reduced Hamiltonian that limited interactions to only nearest-neighbor exchange ($J$) of electrons with antiparallel spin. The Hamiltonian in Ref. \cite{hirschss} is equivalent to the particular choice $V=J$ and $\Delta t=0$ in the more general reduced Hamiltonian considered in this paper; we demonstrate this in App. D.\\

\subsection{One Dimension}

The spin-split state is equivalent to the Rashba-like state in one dimension. In Fig. (7-a), we depict a spin-split state in which the spin is quantized along a direction perpendicular to the electron's momentum; in Fig. (7-b) we show the equivalent Rashba-polarized state. As a result of this equivalence, they share the same dispersion (Eq. (\ref{disp rash 1d})) and ground state energy (Eq. (\ref{ener rash 1d})) with $\R =0$. \\

\begin{figure}[h]
		\subfigure[]{\resizebox{8cm}{!}{\includegraphics[width=7cm, height=3cm]{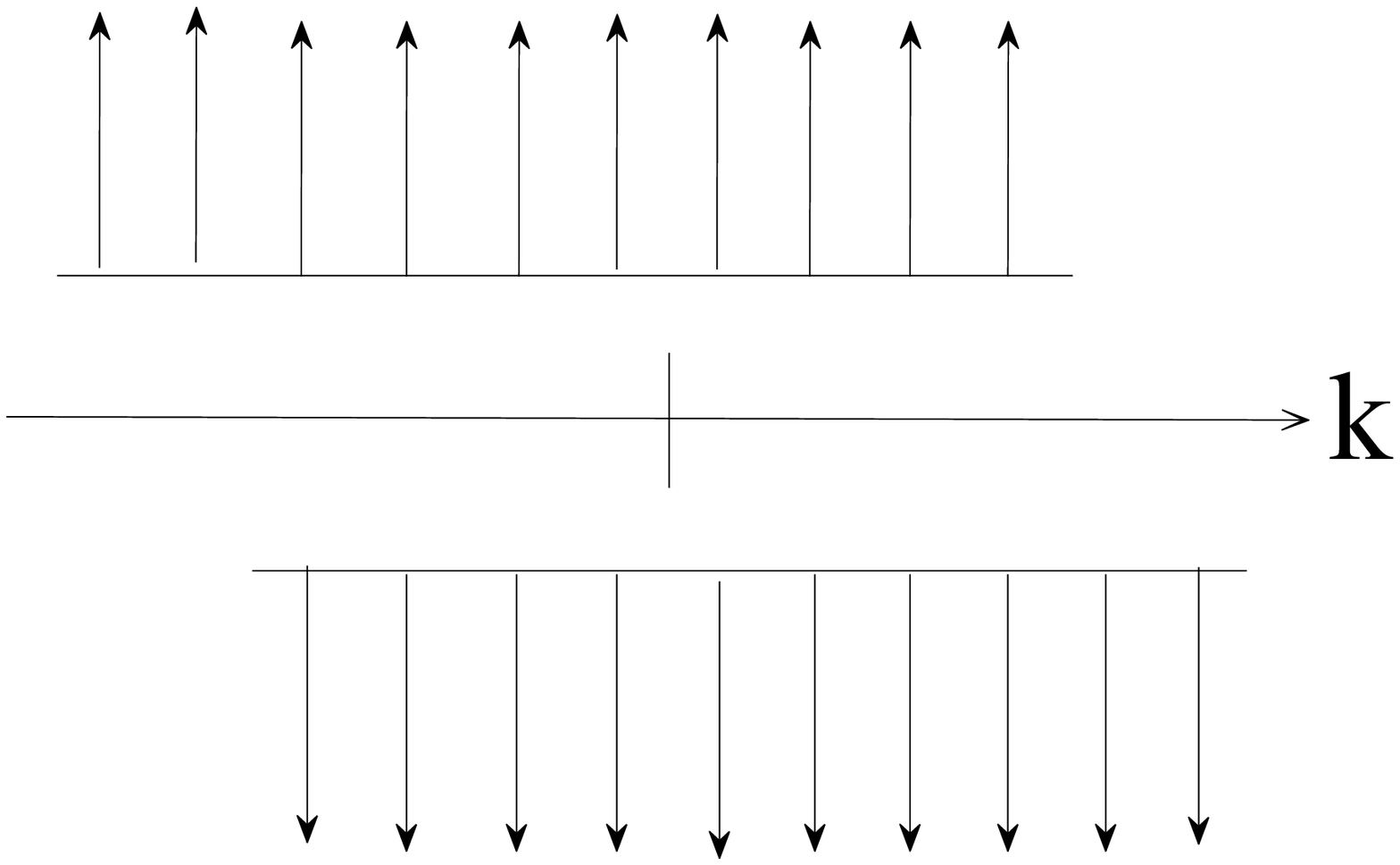}}}
		\subfigure[]{\resizebox{8cm}{!}{\includegraphics[width=7cm, height=3cm]{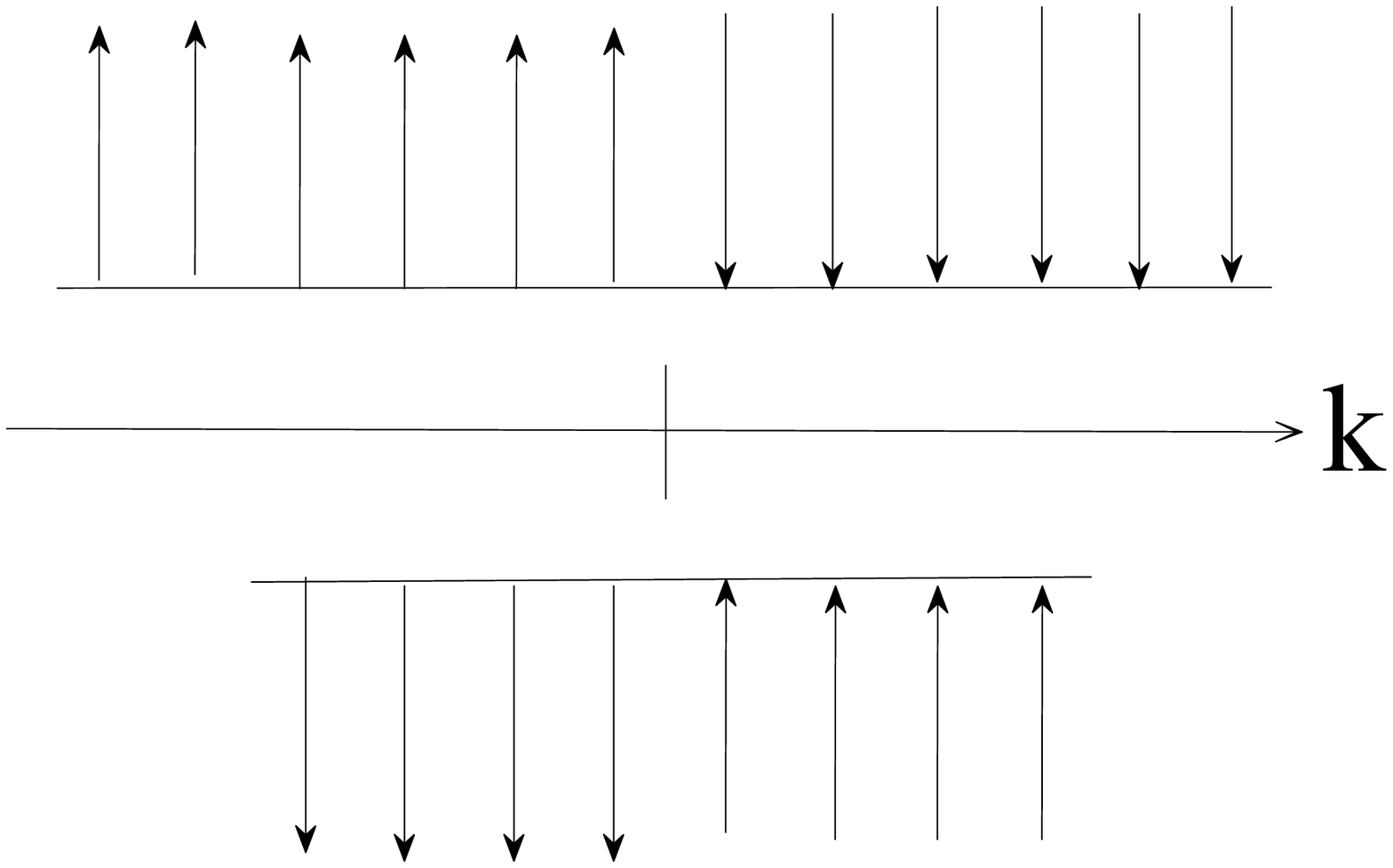}}}
		\caption{(a) \, The spin-split state in one dimension. Spin up and spin down electrons are shown divided by the momentum axis. (b) \, The Rashba state in one dimension. Opposite Rashba polarizations are shown divided by the momentum axis. }
\end{figure}

\subsection{Two Dimensions}

The equivalence between the spin-split and Rashba-like states in one-dimension does not generalize to two dimensions. \\

We begin with the general Hamiltonian in Eq. (\ref{hami}). Since the spin quantization direction is arbitrary in the spin-split state, we choose for convenience that it lies along $\hat{z}$. We decouple the electron-electron interaction with the mean fields
\bal \label{nk ss}
n_{\bfk} =& \;n_{\bfk \up} + n_{\bfk \dow} = \langle c^{\dagger}_{\bfk\up} c_{\bfk\up} \rangle + \langle c^{\dagger}_{\bfk \dow} c_{\bfk \dow}  \rangle \;\;\; \text{and}\\
\Delta_{\bfk} =& \; n_{\bfk\up}  - n_{\bfk \dow} = \langle c^{\dagger}_{\bfk\up} c_{\bfk\up} \rangle  - \langle c^{\dagger}_{\bfk \dow} c_{\bfk \dow} \rangle \label{delt ss}
\end{align} 
The spin-split  state satisfies the symmetry constraints: $n_{\bfk} = n_{-\bfk}, \; \Delta_{\bfk} = - \Delta_{-\bfk}$. Applying these symmetry constraints and the identity $\sum_{\sigma} \big( n_{\bfk \sigma} \; n_{\bfk' \pm\sigma} \big)= (n_{\bfk}n_{\bfk'} \pm \Delta_{\bfk}\Delta_{\bfk'})/2$, we reduce this Hamiltonian and keep only $H_t,H_J^{\scriptscriptstyle{(\bfq = \boldsymbol{0})}}, H_V^{\scriptscriptstyle{(\bfq = \bfk' - \bfk)}}, H_{\Delta t}^{\scriptscriptstyle{(\bfq = \boldsymbol{0})}}$ and $H_{J'}^{\scriptscriptstyle{(\bfq = \boldsymbol{0})}}$.\\

We define geometric factors
\bal
\alpha =& \frac{1}{N} \sum_{\bfk} (\text{cos}\,k_x + \text{cos}\, k_y) \; n_{\bfk}, \label{alph ss} \\
\gamma =& \frac{1}{N} \sum_{\bfk} \big( \text{sin}\,k_x + \text{sin}\,k_y \big) \; \Delta_{\bfk} \label{gamm ss}
\end{align}
and use the convention: $\sigma = +1 (-1)$ for spin up (down). The energy dispersion is
\bal \label{disp ss}
E_{\sigma}(k) = &-2\bigg( \te - \frac{1}{4} (2J- \V) \alpha \bigg) (\text{cos}\,k_x + \text{cos}\, k_y) \notag \\
&- \sigma \bigg( \frac{1}{2} \V \, \gamma \bigg) \big( \text{sin} \,k_x + \text{sin} \, k_y \big).
\end{align}
The ground state energy is
\bal\label{ener ss}
\frac{\langle H \rangle}{N} = -2\te \, \alpha + \frac{1}{2} J \alpha^2 - \frac{1}{4} \V (\alpha^2 + \gamma^2).
\end{align}
The mathematical structure is very similar to that of the Rashba instability (c.f. Eq. (\ref{disp rash}) and (\ref{ener rash})). The difference lies in geometry: (i) $\alpha$ is different for the Rashba and spin-split states because the electrons occupy different regions in momentum space (ii) $\beta$ and $\gamma$ have different geometric weights in their respective integrals.\\  

We observe that the two-dimensional dispersion (Eq. (\ref{disp ss})) decouples into two one-dimensional dispersions as follows \cite{hirschss}:
\bal
E_{\sigma}(k) = &\sum_{\nu} \bigg(-2\big( \te - \frac{1}{4} (2J- \V) \alpha_{\nu} \big) \text{cos}\, k_{\nu} \notag \\
&- \sigma \big( 2 \V \, \gamma_{\nu} \big)  \text{sin} \,k_{\nu}  \bigg), \\
\alpha_{\nu} =& \frac{1}{2N} \sum_{\bfk} \text{cos}\, k_{\nu} \; n_{\bfk}, \\
\gamma_{\nu} =& \frac{1}{2N} \sum_{\bfk} \text{sin}\,k_{\nu} \; \Delta_{\bfk}.
\end{align}
Here the index $v$ runs over the $\{ x,y \}$ directions. If there is an energetic gain to polarizing in the $x$ direction, the symmetry of the lattice permits the same gain for a polarization in the $y$ direction. Hence, there is a spontaneously broken symmetry in which the two spin-differentiated Fermi surfaces are displaced in opposite directions along either x-y diagonals. One example of  the Fermi energy contours of the spin-split state is illustrated in Fig. (1-a). \\

This suggests we may think of the two-dimensional spin-split state as being composed of many one-dimensional systems aligned parallel to the x-y diagonal. For each one-dimensional system, number conservation ensures that the ratio $\alpha_{\nu}^2+\gamma_{\nu}^2$ is a constant upon polarization; this is demonstrated in Eq. (\ref{alph 1d}) and (\ref{beta 1d}). With the identity, $\alpha = 4 \, \alpha_{\nu}, \gamma = 4\, \gamma_{\nu} $ for any $\nu$, we have proven that the geometric combination $\alpha^2+\gamma^2$ is rigorously a constant independent of polarization. An alternative proof is  presented in App. E.\\ 

Minimizing $\langle H \rangle$ while keeping $\alpha^2+\gamma^2$ constant, we obtain for the spin-split state
\bal
\alpha = \frac{2 \, \te}{J}.
\end{align}
The critical $J$ for symmetry breaking is
\bal \label{jc ss}
J_c = \frac{2\,\te}{\alpha_o}.
\end{align}
The dispersion simplifies to
\bal \label{disp ss simp}
E_{\sigma}(k) = &-\frac{\te \,\V}{J} (\text{cos}\,k_x + \text{cos}\, k_y) \notag \\
&- \frac{1}{2} \V \, \gamma \big( \text{sin} \,k_x + \text{sin} \, k_y \big).
\end{align}
Eq. (\ref{nk ss}) to (\ref{disp ss}) and the energy-minimization condition
\bal
\big( E_{\up} (\bfk) \big)_{FS \up} = \big( E_{\dow} (\bfk') \big)_{FS\dow}
\end{align}
may be solved numerically for the unknown factor $\gamma$. Alternatively, we may numerically evaluate $\alpha_o$ for the case of zero polarization and extract $\gamma$ from 
\bal
\alpha_o^2 = \alpha^2 + \gamma^2 = \frac{4 \, \te^2}{J^2} + \gamma^2.
\end{align}
In the limit of small filling, we use a result from App. C to obtain
\bal
\gamma = \frac{2}{\pi} \sqrt{ \bigg( k_{\scriptscriptstyle{F}}\;J_1(k_{\scriptscriptstyle{F}}) \bigg)^2 - \bigg( \pi \frac{\te}{J} \bigg)^2}.
\end{align}

\section{The Role of $V$}

In our previous analysis of the Rashba-like (Sec. III) and spin-split (Sec. IV) instabilities, we have focused on the role of nearest-neighbor exchange $J$ in driving a symmetry breaking but have said little about the role of nearest-neighbor repulsion $V$. In this Section we ask how $V$ (or effectively $\V = V-J'$) affects these two states. We find that $\V$ is pivotal in deciding which of the two ground states are energetically favored - this is discussed in Sec. V-A. In Sec. V-B, we comment on how $\V$ affects the Rashba-like polarization beyond the analytic approximation of Eq. (\ref{anal appr}).\\ 

\subsection{Competitition between the Rashba-like and spin-split instabilities}

The Rashba-like instability with $\R=0$ is energetically compared with the spin-split instability. Except for a small correction that is explained in Sec V-B, they have the same critical $J_c = 2t/\alpha_o$. For convenience, we call this the first critical $J_{c1}$. Which state is preferred for $J>J_{c1}$ is decided largely by a competition between the geometric combinations $\alpha^2 + \beta^2$ and $\alpha^2 + \gamma^2$; the parameter $V$ plays a larger role than parameters $\te$ and $J$. To understand this, consider the difference in energies of the Rashba and spin-split states:
\bal \label{delt h}
\delta \langle H \rangle = \delta \langle H_{\te,J} \rangle + \delta \langle H_{\V} \rangle 
\end{align}
The first term in this difference is $\sim O(\delta \alpha)^2$ because $\langle H_{\te,J} \rangle$ is minimized with respect to $\alpha$ for the spin-split state. The second term is $\sim O(\delta \alpha, \delta \beta)$ and is decisive in determining where the spin-split and Rashba states are degenerate.\\

We first examine the geometric difference between $\beta$ (Eq. (\ref{beta})) and $\gamma$ (Eq. (\ref{gamm ss})). As a first approximation, this difference suffices to explain qualitatively the behavior of $\alpha^2+\beta^2$ relative to $\alpha^2+ \gamma^2$. For any momentum, the geometric weight in the integral of $\beta$ is always less than or equal to the weight in $\gamma$:
\bal \label{weig}
\sqrt{\text{sin}^2\,k_x + \text{sin}^2\,k_y} \le \text{sin}\,k_x + \text{sin}\,k_y.
\end{align}
In one dimension these weights are trivially equal, hence the spin-split and Rashba states are exactly equivalent (see Sec. IV-A). In two dimensions, Eq. (\ref{weig}) naively suggests that $\beta < \gamma$. \\

However, we must consider the factor $\Delta_{\bfk}$ in these integrals. For the same set of parameters, there are generally more electrons polarized in the Rashba state because electrons polarize in all directions, while electrons in the spin-split state polarize only in the x-y diagonal direction. We support this claim numerically in Fig. (8).\\

\begin{figure}[h]
		\resizebox{8.5cm}{!}{\includegraphics[width=8cm, height=8cm]{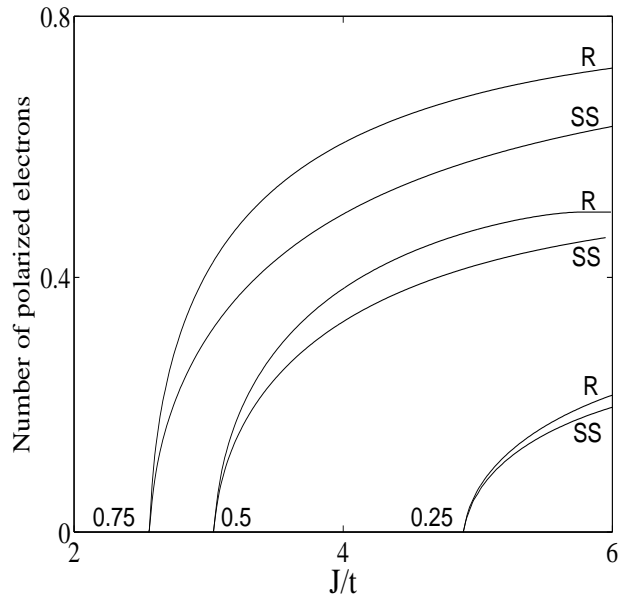}}
	\caption{ The absolute number of polarized electrons vs $J/t$ for the Rashba (labelled $R$) and spin-split (labelled $SS$) instabilities at different band fillings: $\bar{n}=0.25,0.5,0.75$. Parameters: $t=1, \V =3$. For all fillings, the critical $J_{c1}$ for symmetry breaking from the paramagnetic phase is  nearly equal for both Rashba and spin-split states.}
	\label{O vs J for all n}
\end{figure}

\subsubsection{Small filling}

For small filling, the fractional difference in the number of polarized electrons is small - the effect of $\Delta_{\bfk}$ is outweighed by the difference in geometric weights. Hence $\beta < \gamma$. This explains why $\ab$ is a monotonically decreasing function of $J/t$ for small fillings; we show this in Fig. (9-a). Since $\ab<\ag$, a comparison of the energies in Eq. (\ref{ener rash}) and (\ref{ener ss}) implies that the spin-split state is favored for small filling if $\V = V - J'>0$. We demonstrate in Sec. VII that a large positive $\V$ is necessary for either spin-split or Rashba state to exist at all, hence we predict that the spin-split state exists only for small filling.\\ 

\begin{figure}[h]
		\subfigure[]{\resizebox{8.5cm}{!}{\includegraphics[width=8cm, height=7cm]{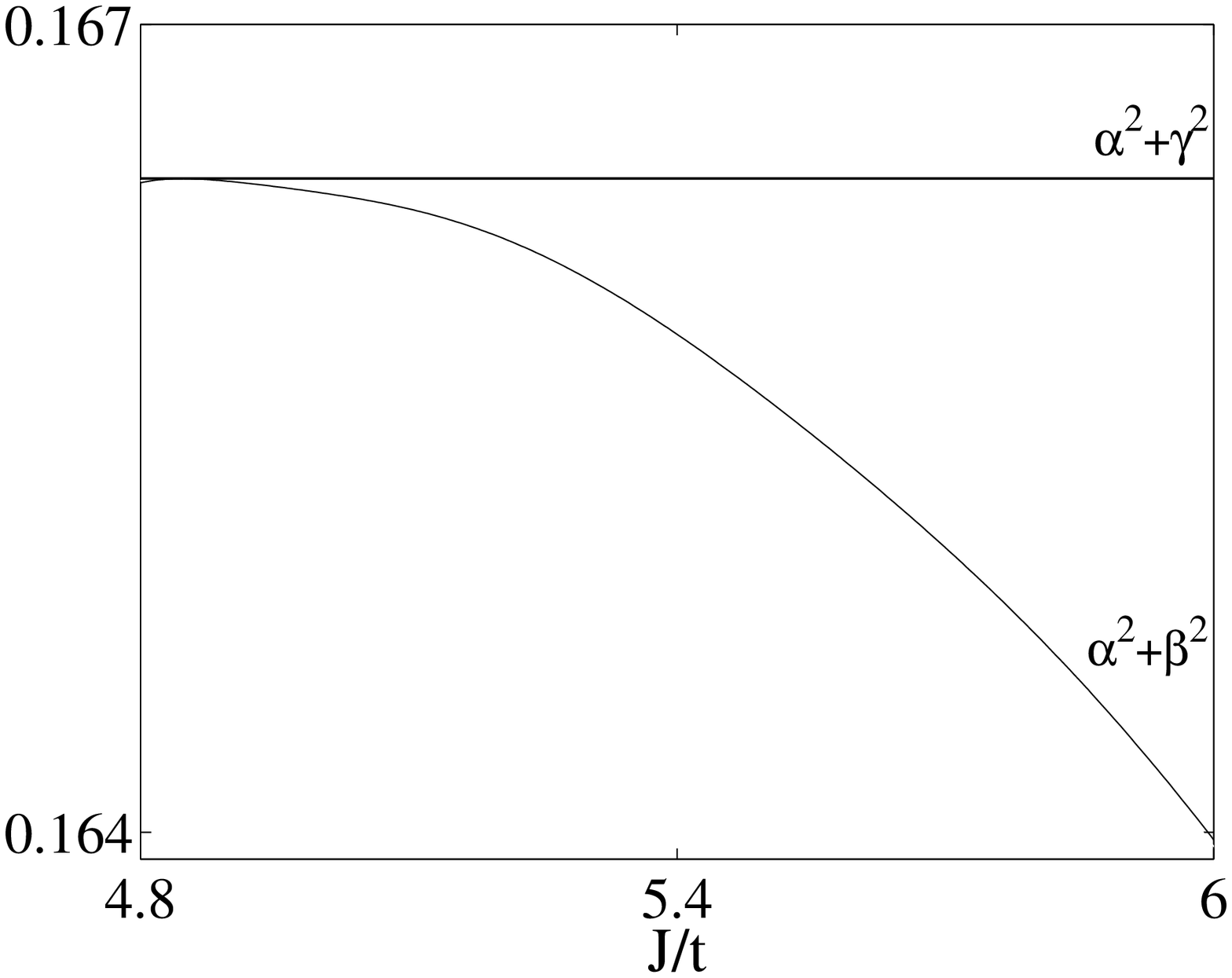}}}
		\subfigure[]{\resizebox{8.5cm}{!}{\includegraphics[width=8cm, height=7cm]{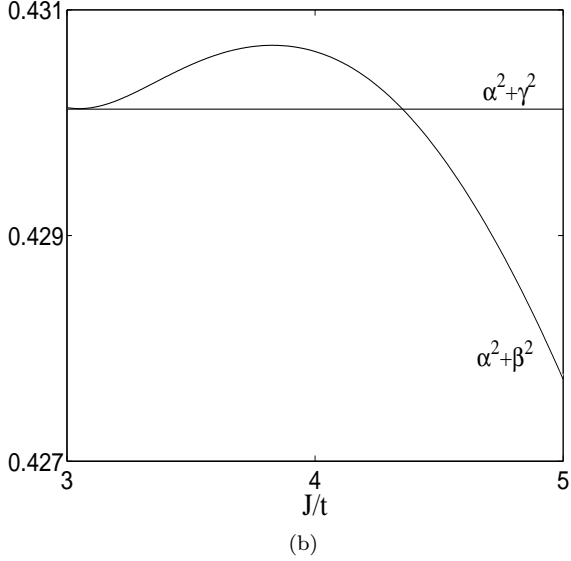}}} 
	\caption{Geometric combinations $\ab$ and $\ag$ vs $J/t$ for (a) $\bar{n}=0.25$ (b) $\bar{n}=0.5$. Parameters: $t=1, \V=3$.}
\end{figure}

\subsubsection{Phase transition above a critical filling}

Above a certain critical filling ($\bar{n}_c = 0.45$), an interesting geometric feature of $\ab$ leads to the possibility of a transition between Rashba and spin-split states at a second critical $J_{c2}$, where these states are degenerate \emph{for a second time}. In Fig. (9-b), we observe that for quarter-filling, $\ab$ initially rises with increasing $J$, then eventually falls and intersects $\alpha^2+\gamma^2$ when $J=J_I$. \\

As we have argued through Eq. (\ref{delt h}), the spin-split and Rashba states are degenerate near this intersection $J_I$. $J_{c2}$ would lie exactly on $J_I$ if not for a small correction $\sim O(\delta \alpha)^2$ due to $\delta \langle H_{\te,J} \rangle$. We recall from Eq. (\ref{alph}) that $\alpha$, the relevant geometric factor in $\langle H_{\te,J} \rangle$, is the negative sum of the kinetic energy; $\alpha$ is smaller for the spin-split state because it polarizes in only one direction and incurs a larger kinetic energy cost. Hence, $\langle H_{\te,J} \rangle_{SS} < \langle H_{\te,J} \rangle_{R}$ and the effect of $\te$ and $J$ through $\delta \langle H_{\te,J} \rangle$ is to reduce $J_{c2}$. \\

We may ask how the magnitude of $\V$ affects $J_{c2}$. In the limit of large $\V$, $\delta \langle H_{\V} \rangle \gg \delta \langle H_{\te,J} \rangle $ hence $J_{c2} \rightarrow J_I$. The limit of $\V \rightarrow 0$ is unphysical for two reasons: (i) we note from Eq. (\ref{disp ss simp}) that the spin-split state has a bandwidth and energy splitting proportional to $\V$ (ii) in Sec. VII we show that both spin-split and Rashba states are unstable against ferromagnetism for small $\V$. In Fig. (10), we plot $J_{c1}, J_{c2}$ and $J_I$ as a function of band filling for $\V=3$. Above the critical filling and for $J_{c1}<J<J_{c2}\approx J_I$, $\ab > \ag$ so the Rashba phase is favored; for $J>J_{c2}$, the spin-split phase is favored. \\

\begin{figure}[h]
		\resizebox{8.5cm}{!}{\includegraphics[width=8cm, height=8cm]{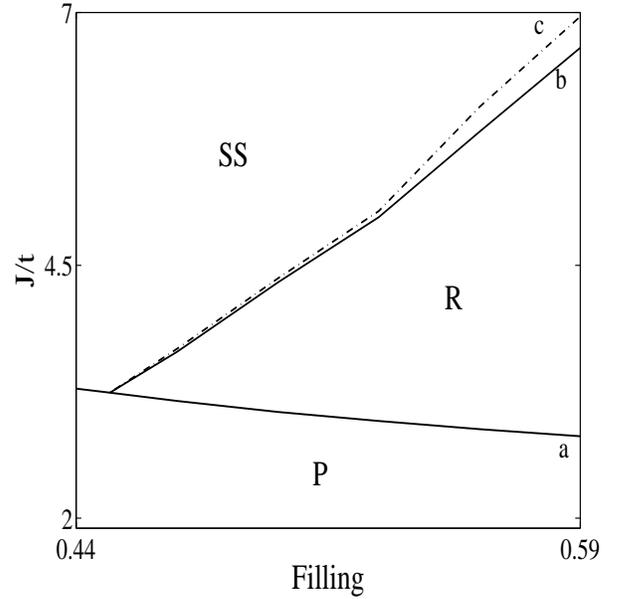}}
	\caption{ Phase diagram for the Rashba (R), spin-split (SS) and paramagnetic (P) states as a function of band filling and $J/t$. $\V =3$. $a$ labels $J_{c1}$; $b$ labels $J_{c2}$; $c$ labels $J_I$, the intersection of the geometric factors $\alpha^2 + \gamma^2$ and $\alpha^2+\beta^2$.}
	\label{jc1 jc2}
\end{figure}

We understand the behavior of $\ab$ and $\ag$ in the following manner: \\

(i) In the limit of large $J/t$, the polarizations of both Rashba and spin-split states saturates to a maximum ($\bar{p} \rightarrow \bar{n}$) and the fractional difference in the number of polarized electrons diminishes. The geometric weights dominate the integrals in $\beta$ and $\gamma$, hence $\beta<\gamma$.\\

(ii) In the limit of small $(J-J_{c1})/t$, the Rashba state has significantly more polarized electrons than the spin-split state, as we may observe in Fig. (8). This allows $\Delta_{\bfk}$ to dominate the integrals in $\beta$ and $\gamma$. The effect of $\Delta_{\bfk}$ becomes more pronounced as $\bar{n} \rightarrow 1$ because the fractional difference in the number of polarized electrons increases with the band filling. Consequently, there exists a critical filling $\bar{n}_c=0.45$ above which $\beta>\gamma$.\\

From (i) and (ii), we deduce that there must be an intersection $J_I>J_{c1}$ where $\ab =\ag$. In addition, (ii) implies that as the band filling is increased, the range of $J/t$ for which $\beta>\gamma$ also increases, hence both $J_I$ and $J_{c2}$ increase. This increase is quite dramatic, as we show in Fig. (10). The spin-split phase is effectively squeezed out of existence by the Rashba phase for large filling because $J_{c2} \gg t$. \\
 
The introduction of the single-particle Rashba interaction ($\R>0$) changes the phase diagram in Fig. (10) in two ways: (i) the spin-split phase will shrink in parameter space (ii) there is no longer a spontaneous broken symmetry for the Rashba state; the paramagnetic phase becomes Rashba-polarized.\\

\subsection{The Role of $\V$ in Rashba-like states}

In our previous discussion of the Rashba-like state in Sec. III-C, we have made an analytic approximation in Eq. (\ref{anal appr}) that is equivalent to claiming that the polarization is independent of $\V$. We have shown that this approximation is only consistent if the resultant polarization is small relative to the band filling. We now go beyond this approximation and ask how $\V$ changes the Rashba polarization.\\

We consider an infinitesimal variation to the ground state energy in Eq. (\ref{ener rash}) due to an increase in polarization
\bal \label{455}
\frac{\delta \langle H \rangle}{N} = \big(J\,\alpha -2\,\te \big)\delta \alpha - \frac{\V}{4} \delta\big(\ab\big).
\end{align}
$\V$ assists polarization if $\delta \big(\ab\big)$ is positive. From our discussion in Sec. V-A, we know that the sign of $\delta \big(\ab \big)$ is always negative for band fillings below the critical level of $0.45$ - we conclude that $\V$ discourages Rashba-like polarization for bands with small filling.\\

For fillings above the critical level, we note from Fig. (9-b) that the slope of $\ab$ levels off at some critical polarization which we call $\bar{p}_c$. $\V$ assists (suppresses) polarization for $\bar{p}<\bar{p}_c \, (\bar{p}>\bar{p}_c)$; in the limit of large $\V$, the polarization will approach this critical value.\\

From Eq. (\ref{455}) we derive the critical $J$ for symmetry breaking
\bal
J_c = J_c^{ss} + \frac{\V}{4\alpha_o} \bigg( \frac{\delta \big( \ab  \big)}{\delta \alpha} \bigg)_o.
\end{align}
We have denoted the critical $J$ of the spin-split state as $J_c^{ss} = 2 \te /\alpha_o$. The second term is a correction to the analytic approximation of Sec. III-C. From our discussion in Sec. V-A, we deduce that $J_c > J_c^{ss}$ ($J_c < J_c^{ss}$) for band fillings less than (more than) 0.45. The ratio of $ \big( \delta ( \ab ) / \delta \alpha \big)_o$ is fixed by the constraint of number conservation and is typically much less than $ \alpha_o$ for Rashba-like systems -  $J_c \approx J_c^{ss}$ for moderate values of $\V$.  \\

\section{General class $J$ ground states}

In this Section we present the formalism for the description of a general Class $J$ ground state. We consider the energetic stability of Class $J$ states other than the spin-split and Rashba-like states and comment on various symmetries such as parity and time-reversal.\\

\subsection{Formalism} 

A general Class $J$ ground state may be characterized by their spin structure $\psi(\bfk)$ (defined in Eq. (\ref{basis})) and polarization structure $\Delta(\bfk)$ (defined in Eq. (\ref{delt})). The various possible ground states are distinguished by the component of the interacting Hamiltonian that describes nearest-neighbor repulsion; for example, see Eq. (\ref{expe hv}). From Eq. (\ref{expe hv}) we may separate the amplitude for nearest-neighbor repulsion
\bal \label{richard}
&\sum_{\delta} V\; e^{i(\bfk -\bfk')\cdot \boldsymbol{\delta}} \notag \\
&= 2\, V \sum_{\nu} \bigg( \; \si k_{\nu} \; \si k'_{\nu} + \; \co k_{\nu} \; \co k'_{\nu} \bigg). 
\end{align}
We consider only the component of Eq. (\ref{expe hv}) that is dependent on $\psi$ and $\Delta$
\bal \label{jimmy}
\frac{\langle H_V^{\scriptscriptstyle{ (\psi,\Delta)}} \rangle}{N} = -\frac{V}{2} \sum_{\nu}& \bigg\{ \bigg( \frac{1}{N} \sum_{\bfk} \Delta_{\bfk} \; \zeta(\bfk) \; \si k_{\nu} \bigg)^2 \notag \\
 +& \bigg( \frac{1}{N} \sum_{\bfk} \Delta_{\bfk} \; \chi(\bfk)\; \si k_{\nu} \bigg)^2 \bigg\}.
\end{align}
In arriving at Eq. (\ref{jimmy}) we have dropped the second term proportional to $\co k_{\nu}$ in Eq. (\ref{richard}); this is necessary to satisfy the symmetry constraint in Eq. (\ref{2nd symm}). \\


We define geometric factors
\bal \label{zeta}
\bar{\zeta}_{\nu} =& \frac{1}{N} \sum_{\bfk} \Delta(\bfk) \; \zeta(\bfk) \; \si k_{\nu} \; \; \; \text{and}  \\
\bar{\chi}_{\nu} =& \frac{1}{N} \sum_{\bfk} \Delta(\bfk) \; \chi(\bfk) \; \si k_{\nu}. \label{chi}
\end{align}
Eq. (\ref{zeta}) and (\ref{chi}) suggest we may interpret $\bar{\zeta}_{\nu}$ and $\bar{\chi}_{\nu}$ as generalized Fourier coefficients of the harmonics $\zeta(\bfk) \, \si k_{\nu}$ and $\chi(\bfk) \, \si k_{\nu}$ respectively. The change in the energy dispersion due to $\zeta$ and $\chi$ is
\bal \label{disp gene}
E_{\Omega}^{(\zeta,\chi)}(\bfk) = -\Omega \;V \sum_{\nu} \si k_{\nu} \; \bigg( \zeta(\bfk) \; \bar{\zeta}_{\nu} + \chi(\bfk) \; \bar{\chi}_{\nu} \bigg).
\end{align}
Including the effects of $J'$, the ground state energy becomes
\bal \label{ener gene}
\frac{\langle H \rangle}{N}  = -2 \te \alpha + \frac{1}{2} J \alpha^2 - \frac{1}{4} \V \bigg( \alpha^2 + 2 \sum_{\nu} \big( \bar{\zeta}_{\nu}^2 + \bar{\chi}_{\nu}^2 \big) \bigg). 
\end{align}

\subsection{Energetic Considerations of other Class $J$ Ground States}

In Sec. III, IV and V we chose to analyze the spin-split and the Rashba-like states because they are the simplest examples of Class $J$ ground states. In this Section, we demonstrate that they are also the most energetically stable phases in their class. The general principle behind these claims is that anisotropy is energetically costly.\\



In Sec. II we learned that the spin polarization as a function of momentum is given by $\psi(\bfk) =$ $ \zeta(\bfk) + i\,\chi(\bfk) = \langle \sigma^x \rangle -i\,\langle \sigma^y \rangle$. For simplicity of notation in this Section, we will omit the implicit dependence on momentum for all functions.
Given a fixed spin structure $\psi$, Eq. (\ref{disp gene}) informs us that $\Delta$ may spontaneously develop a symmetry in any of the four possible harmonics - $\{ \zeta \, \si k_{\nu},\; \chi \, \si k_{\nu} \}$ - that satisfy the symmetry constraint of Eq. (\ref{no magn mome}). 
It is also possible to combine two harmonics to obtain a more favorable ground state. A case in point is the spin-split state. Taking $\psi=1$, we have two possible harmonics - $\si k_x$ and $\si k_y$ - which correspond to a displacement of the Fermi surface in the $\hat{x}$ and $\hat{y}$ direction respectively. These harmonics are complementary in the sense that we may displace the Fermi surface along the $x-y$ diagonal for a greater reduction in energy. Hence, the spin-split state has a polarization with the combined symmetry $\Delta \sim \si k_x + \si k_y$.\\



The spin-split example suggests that a state with the maximum number of complementary harmonics is likely to be energetically favored. We consider a general state that develops a symmetry in all four harmonics; we ask what conditions are necessary to keep all four. We define the reflection operations $R_x$ by ($k_x \rightarrow -k_x,\;k_y\rightarrow k_y$) and $R_y$ by ($k_y \rightarrow -k_y,\;k_x\rightarrow k_x$). For an arbitrary band filling, the condition of no net magnetization (Eq. (\ref{no magn mome})) enforces certain conditions on how $\zeta$ and $\chi$ transform under $R_x$ and $R_y$. The polarized spin $\psi\,\Delta$ has six distinct harmonics  $\{ \zeta^2 \, \si k_{\nu},\; \chi^2 \, \si k_{\nu},\; \zeta \,\chi\, \si k_{\nu} \}$ which must \emph{all} sum to zero to satisfy Eq. (\ref{no magn mome}), i.e. for all momenta $\psi \,\Delta \rightarrow -\psi\,\Delta$ under inversion - there are only two possibilities, which we categorize into subclasses of Class $J$.\\   


The first subclass of ground states is defined to satisfy: (i) $\zeta$ is even under both $R_x$ and $R_y$ (ii) $\chi$ is even under both $R_x$ and $R_y$. As a result, $\Delta(\bfk) = -\Delta(-\bfk)$ - the Fermi surfaces are displaced relative to each other in opposite directions. The choice of $\Theta$ independent of momentum corresponds to the spin-split state.\\


The second subclass is defined to satisfy: (i) $\zeta$ is even under $R_y$ but odd under $R_x$ (ii) $\chi$ is even under $R_x$ but odd under $R_y$. Alternatively, the conditions (i) and (ii) describe an equivalent state if $\zeta$ and $\chi$ interchanged. The combined effect of (i) and (ii) is that the spin polarization rotates $2\pi$ radians around the Fermi contour - it has winding number one. The polarization $\Delta$ has (i) two `s-like' harmonics $\{ \chi \,\si k_x,\;\zeta \, \si k_y\}$ which are even under $R_x$ and $R_y$ (ii) two `d-like' harmonics $\{\zeta \, \si k_y,\;\chi \, \si k_x\}$ which are odd under $R_x$ and $R_y$. Unlike the first subclass, the second subclass has two competing symmetries and will favor one and not the other. We note that both `s-like' and `d-like' instabilities satisfy $\psi \,\Delta \rightarrow -\psi\,\Delta$ under inversion; we elaborate on this in the next Section.\\

The simplest example of the second subclass is $\psi = f_k/|f_k|, \, f_k = \si k_x \pm i \,\si k_y$; this has three distinct harmonics $\{ \si^2 k_x/|f_k|, \;\si^2 k_y/|f_k|,\;\si k_x \,\si k_y /|f_k| \}$. The first two correspond to a relative expansion of one Fermi surface in the $\hat{x}$ and $\hat{y}$ directions; they are complementary in the sense that the Fermi surface may expand in both directions with a combined symmetry $\Delta \sim \sqrt{\si^2 k_x + \si^2 k_y}$ - we call this the Rashba-like phase. The third harmonic corresponds to an expansion along one $x-y$ diagonal and a contraction along the other; this is not complementary with the symmetry $\sqrt{\si^2 k_x + \si^2 k_y}$.\\

This general discussion suggests a great many possibilities among the two subclasses; however, we now argue from energetic considerations that only two really matter. The Rashba-like and the spin-split phases are special because they have the minimum amount of anisotropy needed to reduce the magnetic moment of a polarized state to zero. The Rashba-like phase has the maximum isotropy in charge space and the next-best isotropy in spin space and vice versa for the spin-split phase. In the continuum analogy, the Rashba phase corresponds to an s-wave charge polarization and a p-wave spin polarization, while the spin-split phase corresponds to a p-wave charge polarization and an s-wave spin polarization. \\

A highly anisotropic harmonic $\{\zeta \, \si k_{\nu},$ $\chi \, \si k_{\nu} \}$ encourages an anisotropic polarization that is penalized by high kinetic energy (i.e. greatly reduced $\alpha^2$). In addition, the `surface tension' due to $H_t$ ensures that the overlap between the polarization $\Delta_{\bfk}$ and the harmonic is weak - the generalized Fourier coefficients $\{ \bar{\zeta}_{\nu},\, \bar{\chi}_{\nu} \}$ are diminished. In combination, $\alpha^2 + 2 \sum_{\nu} \big( \bar{\zeta}_{\nu}^2 + \bar{\chi}_{\nu}^2 \big)$ for these anisotropic ground states are generally smaller than that for the spin-split and Rashba-like phases; from Eq. (\ref{ener gene}), we conclude that the Rashba-like and spin-split states are the most favored.\\

We may ask physically why the Rashba-like and the spin-split states are so close in energies. In the spin-split phase, electrons with neighboring momenta have spins in perfect parallel alignment, hence this state benefits from the most reduction in energy $per \;polarized\; electron$ due to the exchange interaction $\V$; this exchange interaction is explained in Sec. II. However, the spin-split state polarizes in one direction and consequently has fewer polarized electrons than the Rashba-like phase. The Rashba phase has a spin structure that rotates $2\pi$ as we go around the Fermi contour; the spins of electrons with neighboring momenta are not perfectly parallel to one another, hence the energy reduction per polarized electron is not as large.  The net effect is that the Rashba-like and spin-split states are almost equally favored by the exchange interaction $\V$. \\


\subsection{Comments on Various Symmetries}

In this Section we ask if Class $J$ grounds states (i) break parity (ii) preserve time-reversal symmetry. We provide a proof for the first and a necessary condition for the second. \\

\subsubsection{Proof of Parity-Breaking}

At least one component of the polarized spin must have a harmonic of the form $\zeta^2 \, \si k_{\nu}$ or $\chi^2 \, \si k_{\nu}$. On inversion, harmonics of this form changes sign. This implies that the spin of the polarized electrons are not invariant under inversion, hence Class $J$ ground states break parity.   \\

\subsubsection{Condition for Time-Reversal Invariance}

For time-reversal invariance to hold, we require that for all momenta $\psi \, \Delta \rightarrow  -\psi \, \Delta$ under parity inversion, i.e. any polarized electron always has a time-reversed partner with opposite momentum and spin polarization. In Sec. VI-B we showed that this condition is satisfied for the first subclass and both `s-like' and `d-like' instabilities in the second subclass; they share the crucial condition that $\zeta\,\chi$ is even under inversion.\\

If we relax this condition, it is still possible to obtain a state with no net magnetization if $\zeta = \zeta(k_x)$ is even under $R_x$ and $\chi=\chi(k_x)$ is odd under $R_x$. $\Delta$ has two competing harmonics: (i) $\zeta(k_x)\,\si k_y$ is `p-like' (ii) $\chi(k_x) \,\si k_y$ is `d-like'. For the favored `p-like' instability, the $\hat{y}$-th component ($-\chi\,\Delta$) of the polarized spin has the symmetry $\si (2\,k_x)\, \si k_y$ which is invariant under inversion, hence time-reversal symmetry is broken. This general discussion suggests that time-reversal breaking states in Class $J$ have at most one harmonic and are not stable relative to multiple-harmonic states like the spin-split and Rashba-like states. The simplest state that breaks time-reversal symmetry is $\psi = e^{-i\,k_x}, \; \zeta\,\chi = \si (2k_x)/2$. \\

\section{Comparison with the Ferromagnetic Instability}

The ferromagnetic instability has a net magnetic moment and is not a Class $J$ ground state. The ferromagnetic ground state has the most isotropic `polarized spin' - it corresponds in the continuum analogy to an s-wave charge and s-wave spin polarization. In this Section, we compare the ferromagnetic with the Class $J$ instabilities. We single out nearest-neighbor repulsion $V$ as the crucial interaction that allows Class $J$ ground states to compete with ferromagnetism.\\ 

We begin with the Hamiltonian in Eq. (\ref{hami}) and decouple the electron-electron interactions with the mean fields defined in Eq. (\ref{nk ss}) and (\ref{delt ss}). In contrast with the spin-split state, the ferromagnetic state satisfies the symmetry constraints: $n_{\bfk} = n_{-\bfk}, \; \Delta_{\bfk} = \Delta_{-\bfk}$. This difference changes the mean-field Hamiltonian significantly. Following a previous work by one of us \cite{amadon}, we keep terms: $H_t,H_{U}^{\scriptscriptstyle{(\bfq = \boldsymbol{0})}},H_J, H_V^{\scriptscriptstyle{(\bfq = \bfk' - \bfk)}}, H_{\Delta t}^{\scriptscriptstyle{(\bfq = \boldsymbol{0})}}$ and $H_{J'}^{\scriptscriptstyle{(\bfq = \boldsymbol{0})}}$.\\

We define the geometric factors
\bal
\alpha =& \frac{1}{N} \sum_{\bfk} (\text{cos}\,k_x + \text{cos}\, k_y) \; n_{\bfk}, \\
\xi =& \frac{1}{N} \sum_{\bfk} (\text{cos}\,k_x + \text{cos}\, k_y) \; \Delta_{\bfk} \; \text{and} \\
\kappa =& \frac{1}{N} \sum_{\bfk} \; \Delta_{\bfk}.
\end{align}

The energy dispersion is 
\bal \label{disp fm}
E_{\bfk \sigma} =& \bigg( -2\,t +2 \Delta t \, \bar{n} + \frac{1}{2}\big( 2J + J' - V \big)\alpha \bigg) \sum_{\nu} \text{cos} \, k_{\nu} \notag \\
                -& \sigma \bigg( \frac{1}{2} \big(J' + V)\xi + 2\Delta t\, \kappa  \bigg) \sum_{\nu} \text{cos} \, k_{\nu} \notag \\
                -& \sigma \bigg(  2\Delta t \, \xi + \frac{1}{2}\big( 4J + U \big) \kappa \bigg) + 2\,\Delta t \, \alpha
\end{align}
and the ground state energy is
\bal
\frac{\langle H \rangle}{N} =& - 2 \big(t - \Delta t \, \bar{n} \big) \alpha + \frac{1}{4} \big( 2J + J' - V \big) \alpha^2 \notag \\
                             & - \frac{1}{4} \big( 4J + U \big) \kappa^2 - \frac{1}{4} \big( J'+V \big) \xi^2 - 2\Delta t \, \kappa \, \xi. 
\end{align}
Unlike the Rashba and the spin-split states, the ferromagnetic state is characterized by a spin-dependent bandwidth and energy baseline. \\

There are more channels through which the ferromagnetic state may gain an energetic advantage. It is no surprise that within the tight-binding Hubbard model, the ferromagnetic instability is favored for a large range of parameters. \\

\subsection{One Dimension}

In Sec. IV, we showed that the Rashba and spin-split states are equivalent in one dimension - for convenience, we call this the \RSS \;state. In this Section, we ask: for what range of parameters is the \RSS \; state favored over the ferromagnetic state in one dimension? \\

We parametrize the ferromagnetic polarization with dimensionless quantity $\bar{\epsilon}$, which is defined: $n_{k \sigma} =1$ for $k \in \{ -k_{\scriptscriptstyle{F}} - \sigma \ep, +k_{\scriptscriptstyle{F}} + \sigma \ep \}$. Minimizing the ground state energy with respect to $\ep$, we obtain an implicit equation for the ground state polarization $\ep^*$:
\bal
& \bigg( 2\pi \, t \, \si \kf  - 4\,\Delta t \, \kf \, \si \kf - 4\,\Delta t \, \co \kf \bigg) \, \si \ep^* \notag \\
=\;& \bigg( U+2J \bigg)\, \ep^* + \bigg( \Delta t \, \co \kf \bigg) \, \ep^* \, \co \ep^* \notag \\
&+ \bigg( 2J \, \text{sin}^2 \kf + V \, \co \kf + J' \bigg) \, \si \ep^*
\end{align}
The limit of vanishing $\ep^*$ determines the critical $\bar{J}_c$ for the onset of ferromagnetism \cite{amadon}
\bal \label{jc ferr 1d}
\bar{J}_c=& \bigg( \pi \, t \, \si \kf  - 2\,\Delta t \big( \kf \, \si \kf +2\,\co \kf \big) \notag \\
& - \frac{U}{2} - V\,\co2\kf -J' \bigg) \bigg/ \bigg( 1+ 2\, \text{sin}^2 \kf \bigg);
\end{align}
from Eq. (\ref{ener rash 1d}) and (\ref{veff}) we extract the critical $J_c$ for the \RSS \;state
\bal
J_c = \frac{\pi \, t - 2\,\Delta t \, \kf}{ 2\,\text{sin}\, k_{\scriptscriptstyle{F}}}.
\end{align}
If $J$ is greater than both $J_c$ and $\bar{J}_c$, ferromagnetism typically wins because it gains an extra energetic advantage through $\Delta t, U$ and $J'$ (c.f. Eq. (\ref{ener rash}),(\ref{ener ss}),(\ref{ener gene})).\\

We analyze the effect of the various Hubbard parameters:\\

(i) In Eq. (\ref{jc ferr 1d}), we note that $V$ is multiplied by $\co 2\kf$ - a large $V$ increases $\bar{J}_c$ in the range $k_{\scriptscriptstyle{F}} \in [\pi/4, 3\pi/4]$. This allows the \emph{RSS} state to compete with the ferromagnetic state. This is illustrated in Fig. (11) for parameters $t=1, \Delta t =0, U=0, J'=0, \R=0$. The \RSS \;state exists in the region where $\bar{J}_c > J>J_c$, which is most likely to occur at half-filling and extends to quarter-filling in the limit of large $V$. For small $V$, $J_c$ never intersects $\bar{J}_c$ (compare curves \emph{a} and \emph{b} in Fig. (11)) and ferromagnetism is always favored. \\

\begin{figure}[h]
		\resizebox{8.5cm}{!}{\includegraphics[width=8cm, height=8cm]{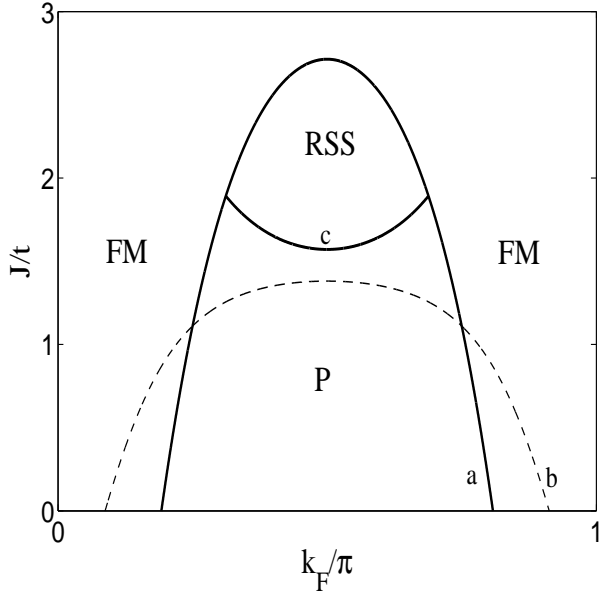}}
	\caption{ Phase diagram for the \RSS, ferromagnetic (FM) and paramagnetic (P) states as a function of the Fermi momentum and $J/t$; one dimension; parameters: $t=1, \Delta t=0, U=0, J' =0$. \emph{a} labels the critical $\bar{J}_c$ for ferromagnetism with $V=5$; \emph{b} labels $\bar{J}_c$ for ferromagnetism with $V=1$; \emph{c} labels the critical $J_c$ for the \RSS \;state.}
	\label{fermi ss}
\end{figure}

(ii) Positive parameters $U$ and $J'$ tend to favor ferromagnetism by reducing $\bar{J}_c$. This narrows the region in which the \RSS \;phase is favored.\\

(iii) The presence of $\Delta t$ breaks the particle-hole symmetry of the phase diagram. It improves (reduces) the stability of the ferromagnetic phase at small (near-full) filling respectively.\\  
  
The introduction of a nonzero $\R$ will Rashba-polarize the paramagnetic phase and shrink the spin-split phase; if $\R \beta$ is comparable to the energetic gain for ferromagnetism, the ferromagnetic phase will diminish as well.\\ 

\subsection{Two Dimensions}

In one dimension, we emphasized the role of nearest-neighbor repulsion $V$ in allowing the Rashba-like and spin-split states to compete with ferromagnetism near half-filling. In this Section we ask if this conclusion generalizes to two dimensions. \\

We proved in Sec. III-B that the energy-minimizing condition is equivalent to equating the single-particle dispersions (Eq. (\ref{disp fm})) of spin up and down electrons at their respective Fermi energies. We define these Fermi energies as $\efu$ and $\efd$ and perform said operation
\bal
&4\,\Delta t\; \xi + \bigg( 4\,J+U\bigg) \kappa \notag \\
= &\bigg( \efu - \efd \bigg) \bigg( 1 - \frac{\Delta t}{t} \bar{n} - \frac{2J+J'-V}{4\,t} \alpha \bigg) \notag \\
 &+  \bigg( \efu + \efd \bigg) \bigg( \frac{J'+V}{4\,t} + \frac{\Delta \, t}{t} \kappa \bigg).
\end{align}
We define the density of states $g(\varepsilon)$. In the limit of zero polarization, we find an expression for the critical $J$
\bal \label{Jc ferr}
&\bigg( 4 \,g(\ef) + \frac{\alpha_o}{2\,t}\bigg) \, \bar{J}_c \notag \\
= \;&1 - U \, g(\ef) + \frac{\Delta t}{t} \bigg( 4 \, \ef \, g(\ef)  - \bar{n} \bigg) \notag \\
&+ \frac{V}{4\,t} \bigg( \alpha_o - \frac{\ef^2\,g(\ef)}{t} \bigg) - \frac{J'}{4\,t} \bigg( \alpha_o + \frac{\ef^2 \, g(\ef)}{t} \bigg).
\end{align}
For a square lattice with only nearest-neighbor hopping, there is a Van Hove singularity at half-filling. Since the singularity
in $g(\varepsilon)$  is logarithmic, the dominant terms in in Eq. (96) the limit $\ef \rightarrow 0$ are
\bal
g(\ef) \bigg( \bar{J}_c + U \bigg) = 1.
\end{align}
As is well-known in a mean-field treatment, ferromagnetism is greatly enhanced where the density of states diverges. In a one-dimensional nearest-neighbor tight-binding model, this divergence occurs for near-empty or near-full bands; in two dimensions the divergence occurs at half-filling. In contrast to one dimension, the two-dimensional Rashba-like and spin-split states cannot compete against ferromagnetism at half-filling. \\ 

We expect that the Rashba-like and spin-split states are stable if a large nearest-neighbor repulsion $V$ suppresses
 ferromagnetism; we know from Eq. (\ref{Jc ferr}) that this only occurs when  the condition $\alpha_o > \ef^2\,g(\ef)/t$
 is satisfied. In one dimension, we showed in Sec. VII-A that this condition restricts the band filling to lie in the range $[1/4,3/4].$ In two dimensions, the density of states is approximately flat away from the divergence at half-filling. We approximate $g(\varepsilon) = 1/(8t)$ and simplify Eq. (\ref{Jc ferr}) to obtain   
\bal
\bar{J}_c =& \bigg( t - 4U - 2\,\Delta t \;\big(2-\bar{n}\big) - \frac{J'}{2}\big(1+\big(1-\bar{n} \big)^2\big) \notag \\
&+ \frac{V}{2}\big( 1-3(1-\bar{n})^2 \big) \bigg) \bigg/ \bigg( 2-\big(1-\bar{n}\big)^2 \bigg).
\end{align}
This result was first derived by one of us \cite{amadon}. Within this approximation, we find that $V$ destabilizes ferromagnetism for band fillings in the range $[0.423,1.577]$; this range is slightly larger than that in one dimension. At half-filling, the approximation breaks down and ferromagnetism
dominates. We have learned in Sec. V-A that the Rashba-like phase likely outcompetes the spin-split state for band fillings larger than $0.45$ or less than $1.55$; this phase diagram is plotted in Fig. (\ref{jc1 jc2}). As we increase the band filling from $0.423$, we predict a transition from the spin-split state to Rashba-like state and then to ferromagnetism near half-filling. A more general treatment than this Section would also allow for antiferromagnetism at half-filling.   \\

\section{Connection with Fermi Liquid $\alpha$ and $\beta$ phases}

Within a tight-binding Hubbard model with on-site and nearest-neighbor interactions, one of us first proposed that electron-electron interactions may spontaneously produce a spin-split phase \cite{hirschss}. In this paper, we demonstrated that the Rashba phase may be produced through a similar mechanism. Wu and coworkers have also explored the spontaneous generation of the $\alpha$ and $\beta$ phases in a Landau-Fermi liquid framework \cite{wu}. The names of these phases should not be confused with the geometric factors defined in Eq. (\ref{alph}) and (\ref{beta}). The spin and momentum-space polarizations of the spin-split and Rashba-like phases are equivalent, in the continuum limit, to that of the $\alpha$ and $\beta$ respectively. In this Section, we compare the formalisms and the conclusions derived from both models.\\

In the Landau Fermi liquid framework, Wu et. al. showed that the $\alpha$ and $\beta$ phases result from a Pomeranchuk instability \cite{pom} in the $l=1$ angular momentum channel and is driven by the phenomenological Landau parameter $F_1^a$. The interaction part of their phenomenological Hamiltonian may be simplified to
\bal
H_{\text{int}} = \frac{1}{2N} \sum_{\bfq} f_1^a(\bfq)  Q^{\mu b}(-\bfq) Q^{\mu b}(\bfq)
\end{align}
with the spin current operator defined as 
\bal
Q^{\mu b}(\bfq) = \sum_{\bfk} c^{\dagger}_{\bfk+\bfq, \alpha} \sigma^{\mu}_{\alpha \beta} \hat{k}^b c_{\bfk \beta}.
\end{align}
The coupling $f_1^a(\bfq)$ is assumed to be a Lorentzian centered around $\bfq=\boldsymbol{0}$. They defined $f_1^a({\bf 0}) = f_1^a$ and decoupled this interaction with a $\bfq = \boldsymbol{0}$ mean field 
\bal
n^{\mu b} = \frac{|f_1^a|}{N} \sum_{\bfk} \langle c^{\dagger}_{\bfk} \sigma^{\mu} \hat{k}^b c_{\bfk}\rangle.
\end{align} 
We define the density of states at the Fermi surface as $N(0)$. The Pomeranchuk instability occurs at $N(0) f_1^a < -2$.\\

The Rashba ground state is achieved by choosing $n^{\mu b}= \bar{n} \, \epsilon^{\mu b}$. The resulting mean-field Hamiltonian is diagonalized by the Rashba basis:
\bal
H_{\text{int}} = \text{sgn}(f_1^a)\bar{n} \sum_{\bfk \eta} \eta \; c^{\dagger}_{\bfk \eta} c_{\bfk \eta}.
\end{align}
This results in an effective spin-orbit field that Zeeman-splits the Rashba energy bands. We compare this phenomenological Hamiltonian with Eq. (\ref{disp rash}) and (\ref{veff}). To cross over from the tight-binding to the Fermi liquid formalism, we limit interactions in the tight-binding model to a thin shell around the Fermi surface and take the continuum limit. By comparing the energy splitting of the Rashba bands in both formalisms, we may identify the phenomenological coupling $f_1^a$ with $-\V \kf^2$ in two dimensions. \\

In the tight-binding Hubbard model, the effective spin-orbit field is due to an exchange interaction in momentum space due to off-site repulsion $V$; there is also a cancelling effect by the pair-hopping $J'$. However, these same processes also increase the bandwith in a manner that suppresses polarization. In the continuum limit, we showed in Sec. V-B that the net effect of $\V$ is to suppress Rashba polarization.\\
 
We reach a similar conclusion in the comparison of the $\alpha$ phase with the spin-split state. The phenomenological model postulates an effective Zeeman field that is necessarily accompanied by an increase in bandwidth within the tight-binding Hubbard approximation. We have shown in Sec. IV that these two effects exactly cancel and that $\V$ neither promotes nor suppresses the spin-split instability. We conclude that the validity of the phenomenological model must lie outside the physical approximations that are assumed in the tight-binding Hubbard model.  
Note that within the Fermi liquid analysis the $\alpha$ phase in two dimensions is predicted to be favored over the
$\beta$ phase when the second derivative of the density of states at the Fermi energy is positive\cite{wu}, which is the case in the band structure
considered here for all band fillings (except half-filled). Instead, in the lattice model we found that  for band fillings above $0.45$ 
the  Rashba phase is favored  (Fig. 10).
\\

We have found that for Class $J$ ground states, $J$ is the greatest driving force for symmetry breaking. It is instructive to translate $H_J^{\scriptscriptstyle{(\bfq = \boldsymbol{0})}}$ to the language of Pomeranchuk instabilities \cite{pom}. In a basis $\Omega$, we decompose an arbitrarily-shaped Fermi surface into its Fourier components
\bal
k_{\scriptscriptstyle{F}}^{\scriptscriptstyle{\Omega}}(\theta) = k_{\scriptscriptstyle{F}}^o + \frac{\delta k_o^{\scriptscriptstyle{\Omega}}}{\sqrt{2}} + \sum_{l>0} \delta k_l^{\scriptscriptstyle{\Omega}} \; \co \big( l\,\theta \big).  
\end{align}
We define $k_{\scriptscriptstyle{F}}^o$ as the Fermi momentum before symmetry breaking. From Eq. (\ref{ht}), the change in kinetic energy due to a deformation of the Fermi surfaces is
\bal
\frac{\delta \langle H_t \rangle}{N} = \frac{1}{4\pi} \;t\; (k_{\scriptscriptstyle{F}}^o)^2 \sum_{l,\Omega} (\delta k_l^{\scriptscriptstyle{\Omega}})^2.
\end{align}
We compare this to the change in the energy of $H_J^{\scriptscriptstyle{(\bfq = \boldsymbol{0})}}$. From Eq. (\ref{expe hj}),
\bal\label{1}
\frac{\delta \langle H_J^{\scriptscriptstyle{(\bfq = \boldsymbol{0})}} \rangle}{N} = -\frac{1}{16\pi^2} \;J\; (k_{\scriptscriptstyle{F}}^o)^4 \sum_{l,\Omega} (\delta k_l^{\scriptscriptstyle{\Omega}})^2.
\end{align}
As we argued in Sec. II, a simplified Hamiltonian with only parameters $t$ and $J$ exhibits symmetry breaking in an arbitrary basis $\Omega$. The critical $J$ is $4\pi \, t/(k_{\scriptscriptstyle{F}}^o)^2$ which agrees with Eq. (\ref{jc rash}) and (\ref{jc ss}) to $O(k_{\scriptscriptstyle{F}}^o)^2$. In addition to the arbitrariness of the basis, Eq. (\ref{1}) also suggests that all Fourier modes $l$ are unstable. We gain specificity when we take into account the effect of the lattice and of $\V$; as we demonstrated in Sec. VI, the ground states with the least anisotropic `polarized spin'- the spin-split and the Rashba-like phases - are the most favored. \\

\section{Discussion}

In this paper we have singled out a class of symmetry-breaking ground states with three properties: (i) electron crystal momentum is a good quantum number (ii) these states have no net magnetic moment (iii) their distribution of `polarized spin' in momentum space breaks the lattice symmetry. Examples of this class of ground states include the spin-split state and the Dresselhaus/Rashba spin-orbit coupled states. Since the Rashba, Dresselhaus and helicity spin-orbit coupled states are degenerate under Coulomb interactions, we just call them Rashba-like. Employing a mean-field approximation, we analyzed these ground states within a tight-binding single-band Hubbard model with all on-site and nearest-neighbor matrix elements of the Coulomb interaction. The relevant matrix elements were found to be nearest-neighbor exchange $J$, pair hopping $J'$ and nearest-neighbor repulsion $V$; $\Delta t$ plays the minor role of reducing the hopping parameter $t$.\\

Because of these symmetry constraints, these ground states lower their energy most effectively through the nearest-neighbor exchange interaction $J$, hence the name Class $J$. Though $J$ has been proposed to lower the \emph{exchange} energy of \emph{localized} electrons with parallel spin\cite{Hei}, the effect of $J$ in \emph{momentum} space is through the \emph{direct} energy\cite{hirschfm}. We have found that $J$ favors a separation of electrons in momentum space, i.e. polarization. This conclusion is independent of the choice of basis, hence $J$ alone does not energetically distinguish between different ground states. In the broken-symmetry phase, a mean-field decoupling of $H_J$ reveals that $J$ lowers the energy by expanding the bandwidth of the single-particle dispersion.\\  

The other relevant interactions are nearest-neighbor repulsion $V$ and pair-hopping $J'$; they act in the combination $\V = V -J'$ to influence both the direct and exchange energies. In the broken-symmetry phase, $\V$ (i) increases the direct energy by compressing the single-particle bandwidth (ii) reduces the exchange energy by splitting the dispersion of electrons with opposite polarizations. We have found that for ground states with highly anisotropic `polarized spin', the advantage gained in exchange energy is outweighed by the cost in direct energy due to $\V$ - these states are energetically unfavorable. Among Class $J$ ground states, the most favored are found to be the spin-split and Rashba-like states - they possess the minimum amount of anisotropy needed to eliminate their net magnetic moment. In the continuum analogy, the Rashba-like phase corresponds to an s-wave charge polarization and a p-wave spin polarization, while the spin-split phase corresponds to a p-wave charge polarization and an s-wave spin polarization.\\

We provided a physical explanation for why the spin-split and Rashba-like states are so close in energy. In the spin-split phase, electrons with neighboring momenta have spins in perfect parallel alignment, hence this state benefits from the most reduction in exchange energy $per \;polarized\; electron$ due to $\V$. However, the spin-split state polarizes in one direction and consequently has fewer polarized electrons than the Rashba-like phase. The Rashba phase has a spin structure that rotates $2\pi$ as we go around the Fermi contour; the spins of electrons with neighboring momenta are not perfectly parallel to one another, hence the exchange energy reduction per polarized electron is not as large. The net effect is that the Rashba-like and spin-split states are almost equally favored by $\V$; they are only distinguished by band filling. \\

In the case of a square lattice, we have found that the spin-split state is favored for small band fillings; above the critical filling of $0.45$, we predict a symmetry breaking from the paramagnetic to the Rashba-like phase at $J_{c1}$ and a second phase transition from the Rashba-like to the spin-split state at $J_{c2}>J_{c1}$. Since all nearest-neighbor interactions on a square lattice except $\Delta \,t$  are particle-hole symmetric, $\V$ produces an effective Zeeman field that is extremized at half-filling. This suggests that differences between the spin-split and Rashba-like states are magnified near half-filling - in particular, the Rashba-like state has significantly more polarized electrons than the spin-split state, hence the exchange energy favors Rashba. For near empty (or full) filling, the difference in polarization is not as pronounced and as a consequence the spin-split state is favored.\\

We have found that a Rashba-polarized system that is conventionally explained by a relativistic single-particle interaction ($\R$) with an electric field is indistinguishable from a system with a smaller (even zero) $\R$ but larger $J$ (and $\V$ for larger band fillings). The effects of $J$ and $\V$ are found to be strongest at half-filling. We have compared our model with the Rashba-polarized surface states in $Au(111)$. Due to its near-empty band, we concluded that electron-electron interactions are too weak to explain the large energy splitting in $Au(111)$. \\

We have compared the energetic stability of Class $J$ ground states with ferromagnetism, an instability with the most isotropic `polarized spin'; in the continuum analogy, ferromagnetism corresponds to an s-wave charge and s-wave spin polarization. Because it has a less restrictive symmetry constraint, we found many more channels through which the ferromagnetic state may gain an energetic advantage. The ferromagnetic instability is favored for a large range of parameters, as is confirmed by the abundance of ferromagnetic materials in nature. The crucial interaction that allows Class $J$ grounds states to compete with ferromagnetism is nearest-neighbor repulsion $V$. The particle-hole symmetric repulsion is found to maximally suppress ferromagnetism at half-filling; in contrast, we have shown that its effects on the direct and exchange energies of Class $J$ states are self-cancelling. Hence we expect that the Rashba-like and spin-split states are stable in systems with large $V$ and only for a certain range of band fillings. From a mean-field treatment on a square lattice, this range is found to be between quarter- to three-quarter-filling in one dimension; in two dimensions we argue that this range is enlarged to $[0.423, 1.577]$ with an exception at half-filling, where there is a Van Hove singularity. \\ 

In this paper our analysis was done on a square lattice assuming only nearest-neighbor hopping terms in the single-particle Hamiltonian. As a result, all electron-electron interactions with the exception of correlated hopping $\Delta t$ are particle-hole symmetric - it is not surprising that the effects of $J$ and $\V$ are maximized at half-filling. The conclusions that are \emph{specific to band filling} will change (i) if next-nearest-neighbor  hopping terms are relevant  or (ii) if we analyze a different lattice (e.g. honeycomb, triangular). In particular, we expect that the regions of stability for the different Class $J$ and ferromagnetic states will shift. For example, if we include next-nearest-neighbor hoppings on the square lattice, the Van Hove singularity moves away from half-filling - this may stabilize the Rashba-like and spin-split states at half-filling. In the three-dimensional generalizations of the spin-split and Rashba-like states, we believe our conclusions with regards to band filling are applicable to simple cubic and BCC lattices. \\

\section{Appendix}

\subsection{Rashba tight-binding approximation}
We formally derive the Rashba tight-binding Hamiltonian discussed in Ref. \cite{mireles}. The free-space Rashba spin-orbit term is
\bal
H_R = \R (\sigma_x k_y - \sigma_y k_x)
\end{align}
for a single electron. We generalize this to a lattice as follows
\bal \label{hr gene}
H_R = \sum_{\bfk \sigma \sigma'} \bigg( Y_{k_y} \cds \big( \sigma_x \big)_{_{\sigma \sigma'}} \cns - X_{k_x} \cds \big( \sigma_y \big)_{_{\sigma \sigma'}} \cns \bigg).
\end{align}
Here we have neglected band indices.\\

We Fourier transform Eq. (\ref{hr gene}) and employ the symmetry constraints (i) $X$ is not a function of $k_y$; $Y$ is not a function of $k_x$ (ii) $X$ and $Y$ are odd functions. Denoting lattice sites as $j$ and $j'$,
\bal
& H_R = \notag \\ 
&\frac{i}{\sqrt{N}} \sum_{\substack{jj' \\ \sigma \sigma'}} \bigg(  \sum_{k_y} Y_{k_y} \si \big(k_y(y_j-y_{j'})\big) \delta_{x_j,x_{j'}} \cdj \big( \sigma_x \big)_{\scriptscriptstyle{\sigma \sigma'}} \cnj \notag \\
&- \sum_{k_x} X_{k_x}  \si \big(k_x(x_j-x_{j'})\big) \delta_{y_j,y_{j'}} \cdj \big( \sigma_y \big)_{\scriptscriptstyle{\sigma \sigma'}} \cnj \bigg).
\end{align}

Symmetry on a square lattice imply $X_u=Y_u$. The nearest-neighbor approximation involves keeping only terms with $x_j-x_{j'}=\pm 1$ and $y_j-y_{j'}=\pm 1$. In combination, these are equivalent to setting $X_u = Y_u = \R \, \si u$. Hence we arrive at Eq. (\ref{rash hami}).\\

\subsection{Why minimize with respect to $\alpha$ and $\beta$}

In Eq. (\ref{defi G}) and (\ref{mini G}), we minimized $G$ with respect to $n_{\bfk  \eta}$ on the respective Fermi surfaces. This is equivalent to minimizing $\langle H \rangle - \gamma (\alpha^2 + \beta^2)$ with respect to $\alpha$ and $\beta$ for the following reasons:\\

(i) Keeping $\alpha^2 + \beta^2$ constant is equivalent to fixing particle number. This is strictly true in one dimension and approximately true in two dimensions for small to medium polarization.\\

(ii) The Hohenberg-Kohn Theorem states that the ground state of a many-body Hamiltonian satisfies
\bal
\rho(\boldsymbol{r}) \; \Longleftrightarrow \; V(\boldsymbol{r})
\end{align}
where $\rho$ is the density of electrons in real space and $V$ is an external potential. Treating $n_{\bfk \eta}$ as a density in momentum space and $E_{\bfk \eta}(\alpha,\beta)$ as the external potential, a generalization of the Hohenberg-Kohn Theorem yields
\bal
n_{\bfk \eta} \; \Longleftrightarrow \; E_{\bfk \eta} \; \Longleftrightarrow \; \alpha,\beta.
\end{align}\\

\subsection{Critical $J$ at small filling}

At $J=J_c$, 
\bal
\alpha^2 + \beta^2 =  \alpha_o^2
\end{align}
In the small filling approximation, the Fermi surfaces are circular, hence
\bal \label{101}
\alpha_o \approx &\frac{1}{\pi^2} \int_0^{k_{\scriptscriptstyle{F}}} dk \, k \int_0^{2\pi} d\theta \, \text{\text{cos}}(k\,\text{\text{cos}}\, \theta) \notag \\
= &\frac{2}{\pi} \int_0^{k_{\scriptscriptstyle{F}}} dk\;k\;J_o(k) \notag \\
= &\frac{2}{\pi} \,k_{\scriptscriptstyle{F}}\;J_1(k_{\scriptscriptstyle{F}}) \notag \\
= &\frac{2}{\pi} \sum_{m=0}^{\infty} \frac{(-1)^m}{4^m (2m+2) (m!)^2} k_{\scriptscriptstyle{F}}^{2m+2}. 
\end{align}
$J_n(x)$ are Bessel functions of the first kind.\\

\subsection{Connection with Reference \cite{hirschss}}

In Ref. \cite{hirschss} the spin-split state was studied with a reduced Hamiltonian that limited interactions to only nearest-neighbor exchange ($J$) of electrons with antiparallel spin. We apply the formalism in this paper and derive an important result from Ref. \cite{hirschss}, namely the one-dimensional dispersion of the spin-split state. We derive this in the Rashba basis.\\

We define operators $\hat{n}_{\eta} = c^{\dagger}_{\eta} c_{\eta},\; \hat{n} = \sum_{\eta}\hat{n}_{\eta}$ and $\hat{\Delta} = \sum_{\eta}\eta \, \hat{n}_{\eta}$. Following Ref. \cite{hirschss}, we only allow interactions in $H_J^{\scriptscriptstyle{(\bfq = \boldsymbol{0})}}$ (Eq. (\ref{hj})) between electrons with anti-parallel spin. We may deduce from Fig. (7) that interactions between electrons with antiparallel spin are equivalent to a sum of (i) interactions between electrons of the same Rashba polarization if both electrons have momenta with the same sign (ii) interactions between electrons of opposite Rashba polarizations if both electrons have momenta with opposite signs. We change Eq. (\ref{hj}) by making the substitution   
\bal
\hat{n}_{\bfk'}\hat{n}_{\bfk} \longrightarrow &\frac{1+\text{\text{sgn}}(k)\text{\text{sgn}}(k')}{2} \big( \hat{n}_{\bfk'+}\hat{n}_{\bfk+}+ \hat{n}_{\bfk'-}\hat{n}_{\bfk-} \big) \notag \\ 
+& \frac{1-\text{\text{sgn}}(k)\text{\text{sgn}}(k')}{2} \big( \hat{n}_{\bfk'+}\hat{n}_{\bfk-} + \hat{n}_{\bfk'-}\hat{n}_{\bfk+} \big) \notag \\
= &\frac{1}{2} \hat{n}_{\bfk'}\hat{n}_{\bfk}  +\frac{\text{\text{sgn}}(k)\text{\text{sgn}}(k')}{2} \hat{\Delta}_{\bfk'}\hat{\Delta}_{\bfk}.
\end{align}
After mean-field decoupling, the \text{sin}gle-particle energy spectrum is
\bal \label{disp ss 1d j}
E_\eta (k) = &-2\left( t - \frac{J}{\pi} \,\text{sin} \,k_{\scriptscriptstyle{F}} \,\text{cos} \, \epsilon \, \right) \text{cos} \,k\,  \notag \\
&+2\,\eta \; \bigg( \frac{J}{\pi} \,\text{sin}\,k_{\scriptscriptstyle{F}} \, \text{sin}\, \epsilon \bigg) \text{sin} \, |k|.
\end{align}
By comparing Eq. (\ref{disp rash 1d}) with Eq. (\ref{disp ss 1d j}), we understand the spin-split Hamiltonian studied in Ref. \cite{hirschss} as the particular choices $V=J$ and $\Delta t=0$ in the more general reduced Hamiltonian considered here.\\

\subsection{Proof that $\ag$ is a constant}

Consider an infinitesimal variation of $\alpha$ (defined in Eq. (\ref{alph ss})) due to a relative displacement of the Fermi surfaces
\bal \label{proof}
\delta \alpha = \frac{1}{N} \sum_{k,\nu,\sigma} \co k_{\nu} \; \delta \, n_{\bfk \sigma}.
\end{align}
Since the shapes of the Fermi surfaces are unchanged for both spin up and down, this is mathematically equivalent to keeping the Fermi surfaces stationary and instead displacing the origin of the Brillouin zone in opposite directions for spin up and down. We parametrize this displacement of the origin for spin $\{ \sigma=\pm 1 \}$ as $\delta \bfk^{\sigma} = \sigma \,\epsilon \,(\hat{\bf x}+\hat{\bf y})$.
We may rewrite Eq. (\ref{proof}) as 
\bal
\delta \alpha = \frac{1}{N} \sum_{k,\nu,\sigma} \bigg( \co \big( k_{\nu} + \sigma \, \epsilon \big) - \co k_{\nu} \bigg) n_{\bfk \sigma}.
\end{align}
To first order in $\epsilon$
\bal
\delta \alpha = -\frac{1}{N} \sum_{k,\nu,\sigma} \si k_{\nu} \; \sigma \; n_{\bfk \sigma} = - \epsilon \; \gamma.
\end{align}
From Eq. (\ref{gamm ss}), an analogous treatment of $\delta \gamma$ reveals
\bal
\delta \gamma = \epsilon \; \alpha.
\end{align}
Together,
\bal
\frac{\delta \alpha}{\delta \gamma} =  - \frac{\gamma}{\alpha}.
\end{align}

\acknowledgments
A. A. is grateful to D. P. Arovas for an illuminating discussion on minimization and for insightful comments, and to C. J. Wu for teaching him about the $\alpha$ and $\beta$ phases. A. A. was supported by a grant from the Committee on Research at UCSD.

\end{document}